\documentclass[superscriptaddress,aps,pra,showpacs,floatfix]{revtex4-2}
\usepackage[utf8]{inputenc}
\usepackage{graphicx,amsmath,amsfonts,amssymb}
\usepackage{color}
\usepackage[colorlinks=true, allcolors={blue}]{hyperref}
\usepackage{graphicx}
\usepackage{epstopdf}
\usepackage{float}
\usepackage{placeins}
\usepackage{soul}
\usepackage{lipsum}
\usepackage{ulem}

\newcommand{\orcid}[1]{\href{https://orcid.org/#1}{\includegraphics[width=7pt]{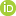}}}

\usepackage[dvipsnames]{xcolor}

\usepackage{soul}%dashword\str{}

\usepackage[T1]{fontenc}
\begin{document}
%\linenumbers
\preprint{APS/123-QED}

\title{Matter-Wave Squeezing from Gouy Phase: Toward a New Tool for Quantum Technology}

\author{Thiago M. S. Oliveira}
%\email{tiago@gmail.com}
\affiliation{Departamento de F\'isica, Universidade Federal do Piau\'i, Campus Ministro Petr\^onio Portela, 64049-550, Teresina, PI, Brazil}

\author{Lucas S. Marinho \orcid{0000-0002-2923-587X}}
%\email{lucas.marinho@ufpi.edu.br}
\affiliation{Departamento de F\'isica, Universidade Federal do Piau\'i, Campus Ministro Petr\^onio Portela, 64049-550, Teresina, PI, Brazil}

\author{F. C. V. de Brito}
\email{crislane.brito@umk.pl}
\affiliation{Institute of Physics, Faculty of Physics, Astronomy and Informatics, Nicolaus Copernicus University in Toruń, ul. Grudziądzka 5, 87-100 Toruń, Poland}

\author{Marcos Sampaio}
%\email{marcos.sampaio@ufabc.edu.br}
\affiliation{Centro de Ci\^{e}ncias Naturais e Humanas, Universidade Federal do ABC,
Avenida dos Estados 5001, 09210-580 Santo Andr\'e, S\~{a}o Paulo, Brazil}

\author{Irismar G. da Paz \orcid{0000-0002-9613-9642}}
\email{irismarpaz@ufpi.edu.br}
\affiliation{Departamento de F\'isica, Universidade Federal do Piau\'i, Campus Ministro Petr\^onio Portela, 64049-550, Teresina, PI, Brazil}

%\date{\today}% It is always \today, today,
             %  but any date may be explicitly specified

\begin{abstract}

	We investigate the Gouy phase emerging from the time evolution of confined matter waves in a harmonic potential. Specifically, we analyze the quantum dynamics of a Gaussian wavepacket that exhibits position--momentum correlations. By tuning the parameters governing its evolution, we reveal intriguing effects, with a particular focus on squeezing. Notably, during the wavepacket evolution quantum spreading and squeezing processes emerge, giving rise to Gouy phase contributions of \(\pi/4\,\mathrm{rad}\), establishing a clear link between the Gouy phase and a purely quantum phenomenon.
	Furthermore, the interplay between wavepacket squeezing and one-dimensional spreading leads to a total Gouy phase accumulation of \(\pi/2\,\mathrm{rad}\) in an oscillation period. Both squeezing and Gouy phase have individually proven valuable in state engineering and quantum metrology. By demonstrating a direct, controllable relationship between these two fundamental processes, our findings expand the realm of quantum-enhanced technologies, including quantum sensing and precision measurement.

\end{abstract}

%\ve{missing: highlight final touch-The feasibility of our proposal, which is ... state-of-the-art, makes it a valuable tool for quantum state engineering.}

%Finally, we analyze the Gouy phase for different confinement %regimes: weak, strong, and resonant. In the resonance, we can find %conditions in which the width of the wavepacket exhibits small %time variations compared to the initial width, resembling a non-%diffracting wavepacket, but still accumulating the total Gouy %phase of $\pi/2$ 

%\keywords{Suggested keywords}%Use showkeys class option if keyword
                              %display ded
\maketitle

%\tableofcontents

\section{Introduction}\label{sec:intro}

The concept of the Gouy phase was first introduced by L. G. Gouy in 1890, and since its initial detection~\cite{gouy1,gouy2}, its properties have been extensively studied~\cite{pang2011phase,pang2011gouy,pang2013manifestation,pang2014wavefront}. 
The Gouy phase has been applied in various fields, such as in the determination of resonant frequencies within laser cavities~\cite{siegman}, phase-matching for strong-field interactions and high-order harmonic generation~\cite{balcou1993phase,lewenstein1995phase,lindner2003high}, and in the analysis the spatial characteristics of high-repetition-rate laser pulses~\cite{Lindner}. 
This phase is also related to the peculiar characteristic of nondiffracting  beams \cite{PabloVaveliuk}, and such beams are a key ingredient for applications ranging from remote sensing \cite{remosense} to biomedical imaging \cite{bio}.
For its role in converting light into vortex beams ~\cite{R5,allen1999iv,guzzinati2013observation,schattschneider2014imaging}, the Gouy phase is a crucial parameter in state engineering with applications in communication and optical tweezing ~\cite{Khoury, polLG}.
In particular, for quantum information protocols, the entanglement provided by correlated photon-pair sources is fundamental, and the Gouy phase has also been investigated in this context \cite{kawase}. 
Recently, the quantum Gouy phase has been measured and confirmed to modify the celebrated photonic de Broglie wavelength,
%showing how it evolves faster for a larger number of photons and a larger mode number. This effect 
expanding the horizon of quantum-enhanced technology for sensitive measurements ~\cite{qgouy,Gu}. 

%Experiences of paticles trapped within the waist of a focused laser beam, which  provides a quadratic potential, offer a promising route to generate highly squeezed states.

	For a particle in a harmonic potential, squeezing leads to a redistribution of uncertainty between position and momentum, thereby creating a correlated state. By enabling state engineering, it modifies the width of the wavepacket, which in turn influences the Gouy phase, offering significant benefits for precision sensing and measurement applications.
	Furthermore, the Gouy phase of photons in squeezed vacuum states has been linked to the notion of topological phases, indicating a geometric component tied to the structure of the state space~\cite{squeezegouy}. Nonlocal effects of this phase have also been documented~\cite{Guo}, where it changes the symmetries connected with self-organization in atomic systems. 
	In the photonic domain, striking a suitable balance between the Gouy phase and the cavity frequency is essential to maintain squeezing in the presence of spatial mode mismatches~\cite{squeezegouy2}. Consequently, by manipulating the geometry of beam propagation or the parameters governing squeezing, one can precisely control the quantum phases \cite{squeezingphase}. This capability is crucial for advancing quantum metrology \cite{metrology, reviewmetrology, atto,gouycompensation}, interferometry \cite{interferometry1, interferometry}, and quantum information processing \cite{Pirandolaqkd, QKD_squeezing_enhancement}.

In multi‑transverse‑mode systems—such as those employing vortex or orbital angular momentum (OAM) beams-squeezing can selectively enhance the signal‑to‑noise ratio of specific modes, thereby improving the overall fidelity of quantum communication. Internal squeezing has additionally been shown to reduce quantum noise and mitigate decoherence~\cite{decoherence}.
The Gouy phase, which varies across transverse modes, offers another degree of control: it can be exploited to ensure that squeezing is applied uniformly across all modes~\cite{polgouy}.
These capabilities are particularly relevant to quantum key distribution (QKD), where such beams have emerged as a promising platform due to their high-capacity encoding of information—a feature that simultaneously presents considerable challenges for practical deployment.  
  The Gouy phase is already considered in QKD protocols using orbital angular momentum (OAM) beams~\cite{GouyOAMKQD1,qkdGouy2,GouyOAMQKD3}, as their spatial‑mode structure induces OAM‑dependent diffraction, resulting in a topological‑charge‑dependent Gouy phase shift. 
  Therefore, the Gouy phase plays a critical role in addressing these challenges, as it enables the monitoring and correction of wavefront perturbations arising from atmospheric propagation, thereby improving the stability and reliability of QKD systems.
  Our work extends these perspectives by explicitly showing how the Gouy phase governs the orientation and manifestation of squeezing, providing a unified framework to jointly control both effects as a potential tool for enhancing the robustness of quantum communication protocols, including QKD.

 Squeezed states of massive particles embody a key
element for ground-based gravitational wave detectors, gravity-induced entanglement, quantum radars, among others \cite{mattersqueeze}.
For instance, cavity-enhanced high-order generated sources are advancing, enabling more accurate measurements and observations at extremely short timescales. Among its potential applications are precise frequency-comb spectroscopy of both electronic and nuclear transitions and low-space-charge attosecond-temporal-resolution photoelectron spectroscopy \cite{atto}.
Effective phase matching is crucial for such sources to operate efficiently and coherently. Conveniently, optimal power enhancement is achieved by the Gouy phase adjustment \cite{gouycompensation}, where it plays a role in fine-tuning the offset frequency for attosecond pulse generation. 
%[https://doi.org/10.1038/s41566-020-00741-3]

%\textcolor{blue}{1-Widely reported in the literature, the Gouy phase is well known for its spatial or temporal confinement origin... 2-}

%[https://doi.org/10.1103/PhysRevLett.123.231108].
% [https://doi.org/10.1364/OE.479583].}.

The relevance of the Gouy phase in coherent matter waves was first demonstrated in \cite{Paz1, Paz2,Arndt} inspiring the subsequent experiments conducted in various systems, including Bose-Einstein condensates \cite{R1, R2, R3} and astigmatic electron waves~\cite{petersen2013measurement}.  We can also mention its rotation effect observed in electron  vortex beams in different situations \cite{guzzinati2013observation,schattschneider2014imaging, R6}.
We have studied this phase in the context of relativistic matter waves \cite{ducharme2015, pazPRL} and its nonlocal effects \cite{poisson}. Digging it further, we found that the Gouy phase for relativistic matter waves contributes to the conversion of orbital angular momentum into spin angular momentum and vice versa \cite{pazPRL}.
Additionally, in the context of position-momentum correlations, have determined the  Gouy phase of SPDC entangled photons \cite{cris1} 
and established its connection to measurements of the photons' quantum correlations.
\cite{cris2}. 
Recently, we examined the influence of the Gouy phase on an interferometric phenomenon using the Cross-Wigner formalism \cite{Marinho2024SciRep}.

In the present work, we investigate the quantum dynamics of a massive particle described by a position-momentum correlated Gaussian wavepacket in a static harmonic potential.
We unravel unique characteristics of matter waves in this setting; 
by examining various frequency relations and correlation parameter, distinct regimes of squeezing and spreading are uncovered. 
We notice that the Gouy phase is directly influenced by the squeezing or expansion of the wavepacket, confirming the relation between this phase and the wavepacket width.
Notably, although the Gouy phase is usually attributed to transverse confinement, the reversal of its concavity under our conditions indicates that it primarily stems from matter-wave squeezing.
In addition, unlike free evolution in one dimension, where the total accumulated Gouy phase is $\pi/4\;\mathrm{rad}$ \cite{Paz1}, the inclusion of squeezing effects leads to a total accumulation of $\pi/2\;\mathrm{rad}$ over one oscillation period.

Our goal is to first establish the fundamental interplay between squeezing and the Gouy phase for a massive particle in a static harmonic potential. Here, the dynamics are governed by fixed frequencies, and the squeezing arises both from the initial position–momentum correlations encoded in the Gaussian wavepacket and from the quantum dynamics away from resonance. This static framework provides a natural starting point for exploring more complex scenarios with time-dependent frequencies.
In realistic settings, decoherence and dissipation can degrade the Gouy-phase–squeezing dynamics. Continuous couplings to stray optical and phonon modes, electromagnetic noise, or unintended trap fluctuations cause damping and phase decoherence, while residual-gas collisions deliver momentum kicks that disrupt correlations and broaden the wavepacket \cite{lossHO,mattersqueeze}. By contrast, intentional parametric modulation of the trap, as in \cite{SutherlandPRL2021}, offers a controlled route to motional squeezing in a time-dependent Hamiltonian. As a potential extension, exploring the Gouy phase in explicitly driven systems could allow its tailoring via time-dependent potentials, offering a versatile platform for controlled squeezing operations. Our ideal, closed-system treatment serves as a clean theoretical baseline for such driven scenarios, and separately, for future generalizations to open quantum systems.

The structure of the work is as follows. In Sec. \ref{sec:II}, we begin by considering that the particle  is described by the Gaussian wavepacket of initial width $\sigma_0=\sqrt{\hbar/(m\omega_0)}$. 
We assume that the initial state is position-momentum correlated. Then, this initial wavepacket evolves under the propagator of the harmonic oscillator with frequency $\omega$. 
Owing to the quadratic nature of the dynamics in both position and momentum, the wavepacket remains Gaussian throughout its evolution.
 We present the analytical expressions for the parameters characterizing the evolved wavepacket, such as the Gouy phase and width, and find that their temporal behaviors are closely connected. 
In Sec. \ref{sec:IIA} we study the frequency and position-momentum correlation effects on the Gouy phase and wavepacket behavior for a fixed value of time.
We investigate the regimes where the current frequency is much smaller ($\omega\ll\omega_0$) and much larger ($\omega\gg\omega_0$) than the fundamental frequency, respectively, in Secs. \ref{sec:IIB} and Sec. \ref{sec:IIC}. We study the dynamics of the Gouy phase and wavepacket in the resonance, in Sec. \ref{sec:IID}, by considering $\omega=\omega_0$.
 Notably, when the position--momentum correlation parameter is zero, the wavepacket width remains constant in time, \(B(t, \gamma=0)=\sigma_0\). Under these conditions, there is neither diffraction nor squeezing, causing the phase to vary linearly with time and to lose the characteristic arctan dependence associated with Gaussian wavepackets.
We further investigate the limit \(|\gamma|\ll 1\) in Sec. \ref{sec:weak_correlation}, in which the wavepacket width exhibits small oscillations around \(\sigma_0\), undergoing minimal spreading and squeezing. Interestingly, the resulting time-dependent phase deviates slightly from the linear profile but preserves the arctan-like Gouy phase signature. 
In Section \ref{sec: Fisher}, we apply the connection between the Gouy phase and matter-wave squeezing as an alternative tool to describe the precision bounds in parameter estimation within the framework of quantum metrology. It establishes a direct conceptual link between Gouy-phase–induced squeezing and optimal parameter estimation, opening a pathway to phase-sensitive quantum control protocols. Section~\ref{sec:conclusion} summarizes our findings and presents final remarks.

\section{Harmonic evolution of a correlated Gaussian wavepacket}\label{sec:II}

It is well known that any mass subjected to a force near a stable equilibrium can, in the leading-order approximation, be modeled as a harmonic oscillator for small vibrations.
In other words, most potentials can be approximated by a parabolic potential in the vicinity of a local minimum \cite{Griffiths2004Introduction}.
Because these phenomena are pervasive in nature, the harmonic oscillator model remains a fundamental cornerstone of modern physics.
The harmonic approximation is applicable across a broad array of physical systems, with ion traps serving as a notable example, since they offer a robust platform for quantum mechanical investigations owing to the extended coherence times of both internal and motional states \cite{WinelandRevModPhys2013}.
In the context of cavity-QED experiments, the relevance of the harmonic oscillator is to describe the electromagnetic field mode inside the cavity, whereas in the ion case, its relevance is associated with describing the ion's motion \cite{WinelandRevModPhys2013}. The applications range from quantum simulations \cite{Blatt2012Nature} to quantum sensing \cite{Cappellaro_2017RevModPhys,Reiter2017NatCommu} and quantum information processing \cite{Cirac2005JournalofPhysicsB,KnightRoyalSocietyofLondon2003}.

In describing confined matter waves, we examine the evolution of a position-momentum correlated Gaussian wavepacket for a massive particle in a harmonic potential.
Although the wavepacket retains its Gaussian form throughout the evolution, its defining parameters—including the Gouy phase and wavepacket width—differ significantly from those associated with free evolution \cite{Paz1}.

We begin by considering the initial state described by the correlated Gaussian wavefunction of transverse width  $\sigma_{0}=\sqrt{\hbar/(m\omega_0)}$ \cite{DODONOV1980PLA}
\begin{equation}
	\psi_0(x') = \frac{1}{\sqrt{\sigma_0 \sqrt{\pi}}} \exp\left[-\frac{x'^2}{2\sigma_0^2} + \frac{i\gamma\, x'^2}{2\sigma_0^2}\right],
	\label{initial_packet1}
\end{equation}
where $\omega_0$ is the intrinsic spread frequency of the initial wavepacket. This state represents an approximately localized state in another potential, which does not necessarily correspond to an eigenstate of the harmonic oscillator, within which we will consider the particle to be confined.

 Initially the study of position-momentum correlations was conducted by Bohm in Ref.~\cite{bohm1951quantum}, since then Gaussian wavepackets exhibiting initial correlations have found widespread applications in various domains, notably in quantum optics \cite{Campos1999JMO}. The position-momentum correlation has been shown to be linked to the Gouy phase~\cite{Paz1}, as well as to the number of interference fringes, squeezing, and quantum coherence effects in the context of the double-slit experiment~\cite{MarinhoPRA2020,lustosa2020irrealism}.
 Additionally, it has been connected to time interference explored through the Cross-Wigner formalism~\cite{Marinho2024SciRep}.
 We can also mention its effect on coherence freezing~\cite{PP2024} and its role in enhancing Gaussian quantum metrology~\cite{Porto2024}. Here, it influences the confined dynamics, leading to distinct modifications in the wavepacket evolution. Therefore, from a practical perspective, the physical origin of the parameter $\gamma$ may be interpreted as arising from the propagation of an atomic beam through a transverse harmonic potential, which effectively acts as a thin lens, thereby imprinting a quadratic phase onto the initial state \cite{JanickeJMO1995}.

The degree of statistical correlation between the position operator $\hat{x}$ and the momentum operator $\hat{p}$ can be obtained through the Person parameter $P=\sigma_{xp}/\sqrt{\sigma_{xx}\sigma_{pp}}$, in terms of which one can express the correlation parameter $\gamma=P/\sqrt{1-P^{2}}$, 
  More generally, one can assess the position-momentum correlations captured by the $\gamma$ parameter through the  covariance matrix of the state.
The dimensionless covariance matrix for a initial correlated Gaussian state can be expressed as follows \cite{RevModPhys2012Lloyd} (refer to Appendix \ref{Wigner} for details):
\begin{equation}\label{eq:iniial_cov_matrix1}
    \boldsymbol{\sigma}(0) =\begin{pmatrix} \;
\frac{\sigma_{xx}}{\sigma_0^2} & \frac{\sigma_{xp}}{\hbar}\\
\frac{\sigma_{xp}}{\hbar} & \frac{\sigma_{pp}}{(\hbar^2/\sigma_0^2)} \;
\end{pmatrix}= \frac{1}{2}\begin{pmatrix} \;
1 & \gamma \\
\gamma & 1+\gamma^2 \;
\end{pmatrix}.
    \end{equation}

The relationship between position-momentum correlations and the phenomenon of squeezing is well known, and we have already addressed it in previous works ~\cite{MarinhoPRA2020,Porto2024}. Thus, consider a single-mode squeezed vacuum state defined as $|S(\zeta)\rangle =  |S(re^{i\phi})|0\rangle$, where $\zeta = r e^{i\phi}$, with $r$ quantifying the degree of squeezing and $\phi$ specifying the orientation of the squeezing ellipse in phase space. The covariance matrix associated with this state is given by~\cite{LandiPRA2024}:
\begin{equation}\label{eq:squeezed_state1}
    \boldsymbol{\sigma_S} = \frac{1}{2}\begin{pmatrix} \;
\cosh 2r - \sinh 2r \cos\phi & - \sinh 2r \sin\phi  \\
- \sinh 2r \sin\phi  & \cosh 2r + \sinh 2r \cos\phi \;
\end{pmatrix}.
\end{equation}

 In squeezed Gaussian states, position and momentum are typically correlated. By comparing the elements of the covariance matrices Eq.\eqref{eq:squeezed_state1} and \eqref{eq:iniial_cov_matrix1}, we can write the correlation parameter explicitly in terms of the squeezing parameters  $\gamma = -\sinh(2r)\sin\phi$. Note that, when $\phi = 0$, the correlation parameter $\gamma = 0$, implying that a nonzero position-momentum correlation arises only when the squeezed state is tilted in phase space. This is evident in the shape of their Wigner functions, which show elliptical contours that are tilted, indicating nonzero covariance. We have calculated the Wigner function for such state, their behavior can be seen in Fig. \ref{FigureWigner} of Appendix \ref{Wigner}. We remind that such correlations may appear in context where the squeezing is absent (Refer to Eq. \eqref{xp_dinamic} in Appendix \ref{Wigner}).

\begin{figure*}[ht]
\centering
\includegraphics[scale=0.22]{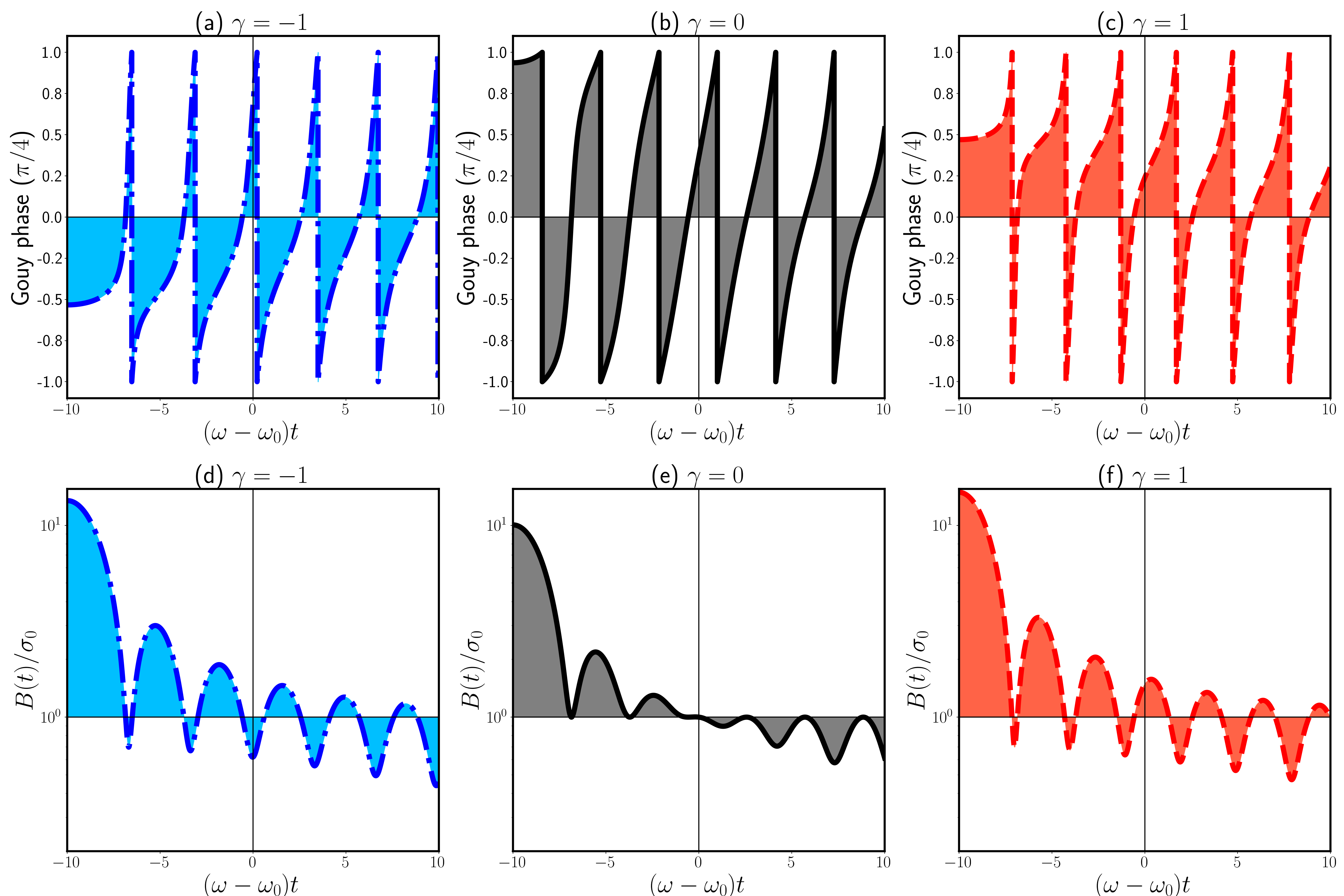}
\caption{Gouy phase and wavepacket width as a function of the natural frequency of the confining oscillator $\omega$, for the intrinsic frequency of the particle $\omega_0=10\;\mathrm{Hz}$ and time evolution $t=1\;\mathrm{s}$. For a qualitative analysis, we have chosen generic values of frequency and time,  only to demonstrate general trends and features of the dynamics. The behavior remains the same as far as the relation between the mismatch frequencies in a real experimental is the same as proposed here. Three values of position-momentum correlations $\gamma$ are considered, with $\gamma =-1$ in (a) and (d) (blue dash-dot),  $\gamma =0$ in (b) and (e) ( black solid line), and $\gamma =1$ in (c) and (f) (red dashed). The wavepacket width oscillates as a function of $\omega-\omega_0$ (bottom plots), presenting regions of squeezing and spread. For $\omega<\omega_0$ and $|\gamma|\neq0$, we observe squeezing and spreading in each period but the spread is more evident. For $\omega>\omega_0$ and $|\gamma|\neq0$, the squeezing is more evident and grows stronger far from the resonance limit where $\omega\gg\omega_0$.  For the position-momentum correlation $\gamma=0$ the wavepacket only spreads for $\omega<\omega_0$ and only squeezes for $\omega>\omega_0$. The squeezing effect is crucial for the Gouy phase accumulation (top plots), which is completely different for the free evolution case, where the Gouy phase is accumulated when the wavepacket spread from the initial width and the total Gouy phase accumulated is only $\pi/4\;\mathrm{rad}$. Here, squeezing and spreading effects contribute to the total acquired Gouy phase of $\pi/2\;\mathrm{rad}$.}\label{Figure1}
\end{figure*}

The temporal evolution of the position-momentum correlated wavepacket within a harmonic potential is obtained as follows
\begin{eqnarray}\label{eq:propagator}
\psi(x,t)&=&\int_{-\infty}^{\infty} dx' G(x,t;x',0)\psi_0(x') , \label{M1}
\end{eqnarray}
 where
\begin{gather}
 G(x,t;x',0)= \sqrt{\frac{m\omega}{2\pi i \hbar \sin\omega t}}\;\;  e^{ \left[\frac{im\omega}{2 \hbar \sin\omega t}[\cos\omega t(x^2+x'^2)-2xx'] \right]}.  
\end{gather}
The kernel $G(x,t;x',0)$ is the nonrelativistic propagator of the harmonic oscillator for a particle of mass $m$ subject to a harmonic potential of frequency $\omega$ \cite{Feynmanbook1965}. Throughout the text, we refer to this frequency as the natural frequency of the confining harmonic oscillator. Finally, $\psi_0(x')$ is the initial wavepacket Eq. \eqref{initial_packet1}.

After some algebraic manipulations in the expression obtained from Eq. (\ref{eq:propagator}), we can write the temporal state of a massive particle with spatial-momentum correlations in a harmonic potential as
\begin{gather}
\psi(x,t)=\frac{1}{\sqrt{B \sqrt{\pi}}}  \exp
\left(-\frac{x^2}{2 B^2} \right)\exp \left(\frac{i m x^2}{2 \hbar R} - i \mu \right),  \label{eq:evolved_state} 
\end{gather}

where
\begin{gather}
B^2(t,\gamma)=\sigma_0^2\Big(\frac{\omega_0}{\omega}\Big)^2\Big\{\sin^2\omega t +[\gamma\sin\omega t+(\omega/\omega_0)\cos\omega t]^2 \Big\}, \label{eq:beam_width_nonressonant}
\end{gather}
\begin{gather}
    R(t,\gamma) = \frac{C(t)\sin\omega t}{\omega C(t)\cos\omega t -(\omega^2/\omega_0)[\gamma\sin\omega t+(\omega/\omega_0)\cos\omega t]}; \\ \nonumber
    C(t)= \sin^2\omega t +[\gamma\sin\omega t+(\omega/\omega_0)\cos\omega t]^2,  
\end{gather}
and
\begin{gather}\label{eq:gouy_nonressonant}
    \mu(t,\gamma) = \frac{1}{2}\arctan \Bigg[\frac{\sin\omega t}{\gamma\sin\omega t +(\omega/\omega_0)\cos\omega t} \Bigg].
\end{gather}
Here, $B(t,\gamma)$ is the wavepacket width,
$R(t,\gamma)$ is the analog of a curvature radius \cite{Paz1}, and $\mu(t,\gamma)$ is the Gouy phase for the propagation through the
harmonic potential. We can also identify $\tau_0=\omega_{0}^{-1}$ as the time scale (``the Rayleigh time"), the analog of Rayleigh length, that governs the spread of the wavepacket in the free evolution and has the same role of the Rayleigh length in the propagation of paraxial light waves \cite{Paz1}.

Similarly to the free evolution, the rate of temporal variation of the Gouy phase is directly related to the wavepacket width by

\begin{equation}\label{eq:GouyRate}
    \frac{\partial \mu (t,\gamma)}{\partial t} =  \frac{\omega_0}{2[B(t,\gamma)/\sigma_0]^2},
\end{equation}
which can be easily obtained by deriving Eq. (\ref{eq:gouy_nonressonant}) with respect to time. As can be seen from Eq. (\ref{eq:GouyRate}), the Gouy phase shift rate can be associated with the wavepacket spatial confinement \cite{feng2001}, with the constant rate characterizing the nondiffracting wavepacket as a direct consequence of its profile spatial invariance \cite{PabloVaveliuk}. From Eq. (\ref{eq:GouyRate}), one obtains that $\frac{\partial^{2}}{\partial t^{2}}\mu(t,\gamma)\;\propto\;\frac{\partial}{\partial t}B(t,\gamma)$, which produces $\frac{\partial^{2}}{\partial t^{2}}\mu(t,\gamma)=0$  for the values of time at which the wavepacket width is either maximum or minimum. Therefore, in this condition, the Gouy phase will change its concavity and/or sign.

In a real experiment, such changes can arise from a variety of noise channels \cite{mattersqueeze,lossHO}. Fluctuations in the trap frequency-caused by electromagnetic field noise, drift in control voltages, or mechanical vibrations-directly perturb $\omega_0$, altering the rate of phase accumulation. Coupling to stray optical modes or phonon modes in the trap structure can lead to gradual heating and damping, modifying $B(t,\gamma)$ over time. Scattering from residual gas molecules imparts discrete momentum kicks, which broaden the wavepacket and reduce the visibility of Gouy-phase signatures. Unintended time-dependent variations in the confining potential can mimic or interfere with the controlled modulation of $B(t,\gamma)$, while preparation noise in position-momentum correlations shifts the initial conditions for phase evolution. Moreover, real-time attempts of continuous measurement would introduce measurement backaction, perturbing the position–momentum correlations that underlie Gouy-phase-induced squeezing.

The Gouy phase in our model is directly connected to the temporal evolution of the wavepacket width through Eq. (\ref{eq:GouyRate}),
which shows that any process altering $B(t,\gamma)$ will also modify the phase evolution. 
However, a rigorous definition of the Gouy phase in open quantum systems, especially within the density matrix formalism, remains an open problem, as decoherence tends to suppress phase-related features. 
Possible approaches to this challenge include interferometric techniques, as discussed in previous works~\cite{VisserJOSAA2012,poisson}, and the analysis of wavepacket spreading in lossy Gaussian channels of transmissivity $\eta$~\cite{RevModPhysLloyd2012}. In the latter case, one may conjecture that relation \eqref{eq:GouyRate} also holds for mixed states as previously proposed in~\cite{Paz1}.
Therefore, our findings for the ideal closed-system provide a clean theoretical baseline for extensions to open quantum systems or to explicitly driven traps with time-dependent potentials that we intend to carry out in the future.

\subsection{Oscillator Frequency and Correlation Effects}\label{sec:IIA}

Here, we analyze the relation between the Gouy phase and the wavepacket width by fixing the propagation time and the initial wavepacket frequency, while varying the natural frequency of the oscillator and the initial position-momentum correlations. Note that the initial frequency of the particle differs from the natural frequency of the harmonic oscillator in which it is confined. As a result, the system is not initially in an eigenstate, leading to an adjustment of the wavepacket shape.
In Fig.~\ref{Figure1}, we plot the Gouy phase and the wavepacket width as a function of the frequency mismatch (\(\omega - \omega_0\)), for different values of the position-momentum correlation parameter \(\gamma\). 

For a qualitative analysis, let us choose generic values for frequency and time,  only to demonstrate general trends and features of the dynamics.
We fix the characteristic frequency of the particle at \(\omega_0 = 10\;\mathrm{Hz}\), the propagation time at \(t = 1\;\mathrm{s}\), and consider three different values of the correlation parameter:  
\begin{itemize}
\item \(\gamma = -1\) (blue dash-dot) in Fig.~\ref{Figure1}(a) and Fig.~\ref{Figure1}(d),  
\item  \(\gamma = 0\) (black solid) in Fig.~\ref{Figure1}(b) and Fig.~\ref{Figure1}(e),  
\item  \(\gamma = 1\) (red dashed) in Fig.~\ref{Figure1}(c) and Fig.~\ref{Figure1}(f).  
\end{itemize}
For clarity, the Gouy phase (top plots) is expressed in units of \(\pi/4\;\mathrm{rad}\), while the wavepacket width \(B(t,\gamma)\) (bottom plots) is given in units of \(\sigma_0\) (the initial wavepacket width). 

As expected from Eq.~\eqref{eq:GouyRate}, the Gouy phase changes its concavity at the time values where the wavepacket width reaches its maximum. Notably, for \(\gamma < 0\), the concavity change occurs in the negative portion of the Gouy phase, whereas for \(\gamma > 0\), the transition appears in its positive portion. This behavior encodes a distinct signature in the Gouy phase, reflecting the change in squeezing rotation direction caused by the sign reversal of the correlation parameter. For \(\gamma = 0\), the concavity transition occurs precisely when the Gouy phase is zero (for the relation between position-momentum correlation and squeezing refer to Fig. \ref{FigureWigner} in the Appendix \ref{Wigner}).

The wavepacket width oscillates as a function of $\omega-\omega_0$ (bottom plots), presenting regions of squeezing and spread. For $\omega<\omega_0$ and $|\gamma|\neq0$, we observe squeezing and spreading in each period, but the spreading is more evident and becomes even stronger far from the resonance regime, where the oscillator's eigenfrequency is much smaller than the particle's intrinsic frequency $\omega\ll\omega_0$. 
On the other hand, for $\omega>\omega_0$ and $|\gamma|\neq0$, the squeezing is more evident and grows stronger far from the resonance limit, where the natural frequency of the oscillator is much larger than the frequency of the initial wavepacket $\omega\gg\omega_0$.  Interestingly, for the position-momentum correlation $\gamma=0$ the wavepacket only spreads for $\omega<\omega_0$ and only squeezes for $\omega>\omega_0$, with the intensity of these effects always smaller than for the cases where $\gamma\ne 0$. Such effects are further enhanced when larger values of $|\gamma|$ are considered, i.e.,  the role of the position-momentum correlations is to produce squeezing and spreading in both intervals of the frequency difference. It is worth  mentioning that squeezing means that the wavepacket width becomes smaller than $\sigma_0$, the width of the initial Gaussian state. This is equivalent to say that the uncertainty in position becomes smaller than that of the vacuum state of the harmonic oscillator while the other quadrature expands.

For matter waves undergoing free evolution, the Gouy phase begins to accumulate as wavepacket spreads from its initial width, reaching a total accumulation of $\pi/4\;\mathrm{rad}$ \cite{Paz1}. In contrast, we notice that for matter waves confined within a one-dimensional harmonic potential, the total Gouy phase acquired is $\pi/2$. 
 Interestingly, the squeezing effect-determined by the frequency difference $\omega-\omega_0$ and position-momentum correlation $\gamma$-plays a crucial role in the Gouy phase accumulation (as shown in the top plots),  connecting the Gouy phase with a purely quantum feature. Therefore, both squeezing and spreading effects contribute to the total Gouy phase of $\pi/2\;\mathrm{rad}$. 
 Note that a total Gouy phase of $\pi/2$ can be acquired in regions without squeezing, as illustrated on the left side of Fig.~\ref{Figure1}(b) and (e), we have an analogous of the focusing effect where half of the phase accumulates as the wavepacket at a maximum width converges to the initial width, while the other half is gained as it expands back to its maximum width. A summary of the frequency and correlation regimes and their consequences on the wavepacket width and Gouy phase is presented in Table I.

In the next subsections, we will analyze the behaviors of the Gouy phase and wavepacket width for different regimes of the harmonic oscillator's natural frequency by varying the propagation time and the initial position-momentum correlations.

\begin{table}[H]
	\centering
	\caption{Summary of wavepacket width and Gouy phase behavior as a function of frequency mismatch \(\omega-\omega_0\) and position-momentum correlation \(\gamma\). }. 
	\label{tab:results}
	\begin{tabular}{|c|c|p{6cm}|p{6cm}|}
		\hline
		\textbf{Frequency Regime} & \textbf{\(\gamma\) Condition} & \textbf{Wavepacket Width Behavior} & \textbf{Gouy Phase Accumulation} \\
		\hline
		\(\omega < \omega_0\) & \(|\gamma|\neq 0\)  & Oscillates with both squeezing and spreading; spreading dominates, especially when \(\omega\ll\omega_0\). & Total Gouy phase of \(\pi/2\,\mathrm{rad}\) (squeezing contribution). \\
		\hline
		\(\omega > \omega_0\) & \(|\gamma|\neq 0\)  & Oscillates with both squeezing and spreading; squeezing dominates, especially when \(\omega\gg\omega_0\). & Total Gouy phase of \(\pi/2\,\mathrm{rad}\) (squeezing contribution). \\
		\hline
        \(\omega > \omega_0\) & \(\gamma=0\)       & Only squeezing is observed; effects are less intense than for \(|\gamma|\neq 0\). & Total Gouy phase of  \(\pi/2\,\mathrm{rad}\) (squeezing contribution). \\
		\hline
		\(\omega < \omega_0\) & \(\gamma=0\)   &  Focusing effect; Spreading and contraction are observed; effects are less intense than for \(|\gamma|\neq 0\). & Total Gouy phase of \(\pi/2\,\mathrm{rad}\) (no squeezing contribution). \\
		\hline
	
	\end{tabular}
\end{table}

\subsection{Time and Correlation Effects for $\omega \ll \omega_0$}\label{sec:IIB}

The Gouy phase and wavepacket features are further investigated here as a function of time in the far-from-resonance regime, considering the case where the natural  frequency of the harmonic oscillator is much smaller than the intrinsic frequency of the particle, i.e., \(\omega \ll \omega_0\).

\begin{figure*}[!ht]
\centering
\includegraphics[scale=0.28]{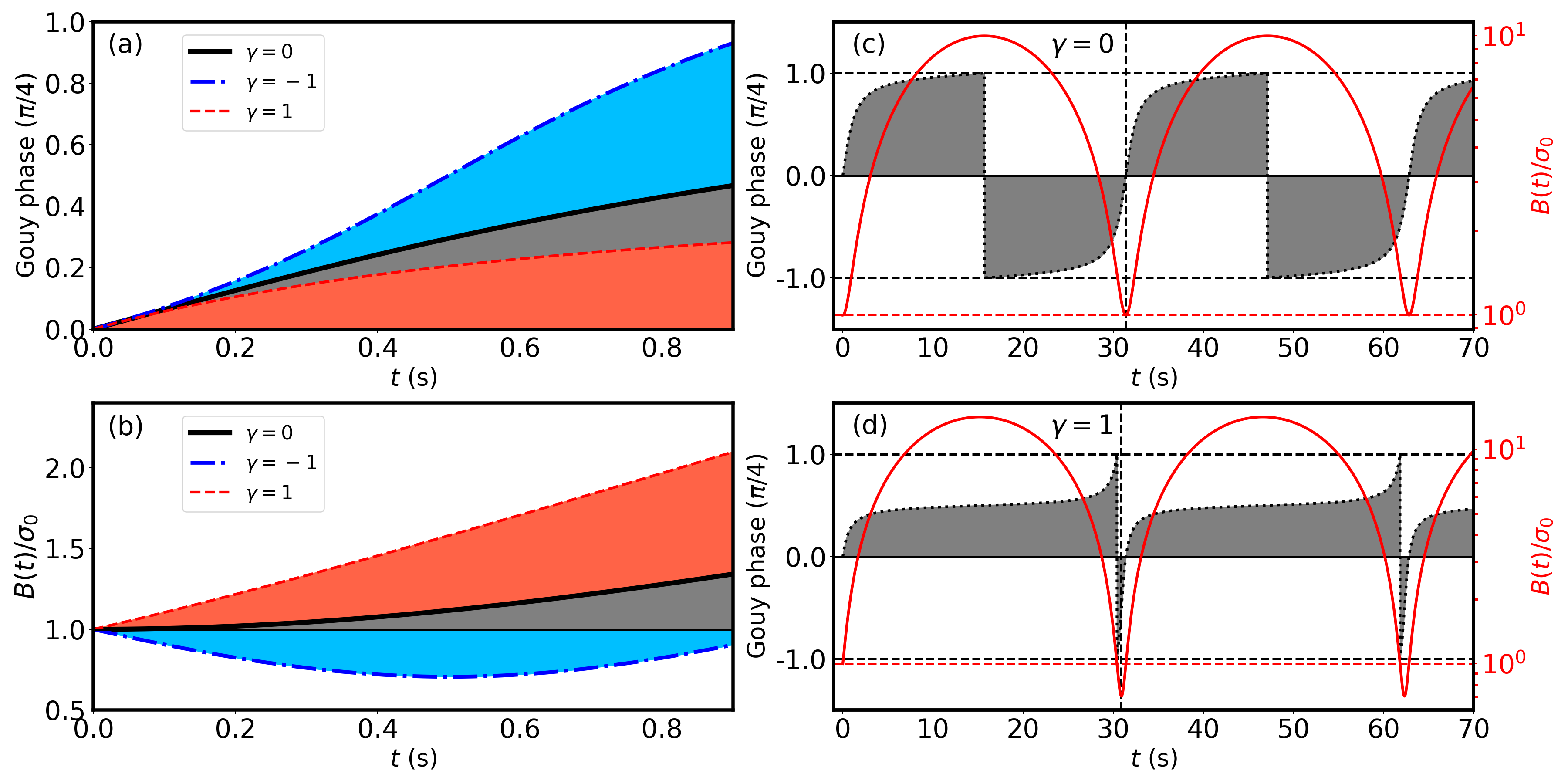}
\caption{Temporal behavior of the Gouy phase and wavepacket width for low harmonic oscillator's frequency $\omega \ll \omega_0$. We consider  $\omega = 0.1 \; \text{Hz} \ll \omega_0 = 1 \; \text{Hz}$ (values chosen arbitrarily only to show the tendency of the dynamics) and three values of the correlation parameter $\gamma$. (a) Gouy phase and (b) wavepacket width for small values of the propagation time. In (c) and (d) Gouy phase and wavepacket width at the same plot for $\gamma=0$ and $\gamma=1$, respectively. Here, large values of propagation time are considered. In (c) and (d) the Gouy phase corresponds to the left scale and the black color while the wavepacket width corresponds to the right scale and the red color. The free dynamics are recovered for smaller values of time where the  Gouy phase evolves to the maximum value of $\pi/4\;\mathrm{rad}$. Within this range of time and for $\gamma=0$ and $\gamma=1$, the wavepacket only spreads, which implies a slow evolution of the corresponding Gouy phase to its maximum value $\pi/4\;\mathrm{rad}$. For position-momentum anti-correlation $\gamma=-1$, the wavepacket first squeezes from the initial width to a minimum width and then spreads from it. Both effects contribute to the Gouy phase, which evolves rapidly to the maximum value $\pi/4\;\mathrm{rad}$ for $\gamma=-1$. For larger values of propagation time, the wavepacket width presents local maxima and minima values. These values of times for the maxima and minima wavepacket width are related with changes in the concavity and/or sign of the corresponding Gouy phase. For $\gamma=0$, within the range of frequencies considered here, the wavepacket never squeezes.  We see an effect similar to focusing (lens effect) on one dimension when the wavepacket propagates from its maximum width at around $t=15.7\;\mathrm{s}$ to the next maximum at around $t=47.1\;\mathrm{s}$.  Within this time interval the total Gouy phase accumulated is $\pi/2\;\mathrm{rad}$. For $\gamma=1$ the squeezing of the wavepacket is crucial for the total accumulation of $\pi/2\;\mathrm{rad}$ for the Gouy phase. }\label{Figure2}
\end{figure*}

The behavior of the Gouy phase is directly linked to the wavepacket width, as indicated by Eq.~\eqref{eq:GouyRate}. 
We confirm the validity of this relation over time by examining Fig.~\ref{Figure2}, where the Gouy phase (in units of \(\pi/4\;\mathrm{rad}\)) and the wavepacket width (in units of \(\sigma_0\)) are plotted against time in the regime \(\omega \ll \omega_0\). 
We consider a natural frequency of the oscillator, \(\omega=0.1\;\mathrm{Hz}\), an intrinsic frequency associated with the initial wavepacket of the particle, \(\omega_0=1\;\mathrm{Hz}\), and three different values of the correlation parameter \(\gamma\).
In Fig.~\ref{Figure2}(a), we show the Gouy phase, and in Fig.~\ref{Figure2}(b), the wavepacket width at shorter timescales. In Fig.~\ref{Figure2}(c) and Fig.~\ref{Figure2}(d), we extend the evolution to longer times and display both the Gouy phase and the wavepacket width on the same plot for \(\gamma=0\) and \(\gamma=1\), respectively. 
In Fig.~\ref{Figure2}(c) and Fig.~\ref{Figure2}(d), the Gouy phase is represented in black, associated with the left axis, while the wavepacket width is plotted in red,  aligned with the right axis.

 As observed, for short times, the free-dynamics behavior is recovered, with the Gouy phase evolving to its maximum value of \(\pi/4\;\mathrm{rad}\). Within this time range, for both \(\gamma=0\) and \(\gamma=1\), the wavepacket  spreads. This spreading occurs more rapidly for \(\gamma=1\), causing a delayed evolution of the Gouy phase towards \(\pi/4\;\mathrm{rad}\) relative to \(\gamma=0\). For position-momentum anticorrelation \(\gamma=-1\), the wavepacket initially contracts to a minimum width before expanding. Both effects contribute to the Gouy phase, which rises quickly to the maximum value of \(\pi/4\;\mathrm{rad}\) for \(\gamma=-1\). The behavior for short times can be observed in the approximate formulas (\ref{eq:B_low_frequency}) (wavepacket width) and (\ref{eq:gouy_low_frequency}) (Gouy phase) in Appendix \ref{ap:Gouy_expressions}. In fact, the first terms of each expression correspond to the free-dynamics expressions \cite{lustosa2020irrealism}.

 Meanwhile, for longer propagation times, the wavepacket width exhibits local maxima and minima points. As indicated by Eq.~\eqref{eq:GouyRate}, the time instants corresponding to the local extrema of the wavepacket width are related to changes in concavity and sign of the associated Gouy phase. For instance, for \(\gamma=0\), as shown in Fig.~\ref{Figure2}(c), the time at which the wavepacket width reaches its maximum corresponds to the instant at which the Gouy phase exhibits a discontinuous sign change, while the time at which the wavepacket width reaches its minimum corresponds to an instant when the Gouy phase undergoes a continuous transition in both sign and concavity.  
 
 Note that for \(\gamma=0\), within the examined frequency range, the wavepacket remains unsqueezed, so that \(B(t,\gamma=0) \ge \sigma_0\). Another interesting characteristic is observed when the wavepacket evolves from its maximum width at around \(t=15.7\;\mathrm{s}\) to the next maximum at around \(t=47.1\;\mathrm{s}\); we observe a one-dimensional focusing effect. In this interval, the wavepacket converges to the minimum value \(\sigma_0\)  and then expands thereafter. Within the time interval where the focusing effect is observed, the total Gouy phase accumulated is $\pi/2\;\mathrm{rad}$.

 For $\gamma=1$, Fig. \ref{Figure2}(d), the maxima and minima values of the wavepacket width remain at the same values of time for $\gamma=0$, i.e., the value of $\gamma$ does not contribute to the time of maxima and minima of the wavepacket width. In contrast to the case $\gamma=0$,
 the times corresponding to the maxima of the wavepacket width are associated with changes in the concavity of the Gouy phase, while the times corresponding to the minima are linked to discontinuous changes in the sign of the Gouy phase.
 Here, we can also observe squeezing of the wavepacket and this effect is crucial for the total accumulation of $\pi/2\;\mathrm{rad}$ for the Gouy phase. Observe that around \(t=0\) we do not have squeezing and the total Gouy phase accumulated from \(t=0\) to \(t=31.4\;\mathrm{s}\) is only \(\pi/4\;\mathrm{rad}\). In contrast, the evolution from \(t=31.4\;\mathrm{s}\) to \(62.8\;\mathrm{s}\) accumulates a total Gouy phase of \(\pi/2\;\mathrm{rad}\) because we have squeezing around \(t=31.4\;\mathrm{s}\) as well as around \(t=62.8\;\mathrm{s}\). We do not present the case $\gamma=-1$ for longer time values, as it would result in a qualitatively analogous case of $\gamma=1$, which we have already presented. An overview of the results discussed in this section is presented in Table II.

\begin{table}[ht]
\centering
\caption{Summary of wavepacket width $B(t,\gamma)$ and Gouy phase $\mu(t,\gamma)$ for $\omega=0.1\,\mathrm{Hz} \ll \omega_0=1\,\mathrm{Hz}$ and three values of $\gamma$. 
Short-time results (up to $t\approx 1\,\mathrm{s}$) match panels (a) and (b) of Fig.~\ref{Figure2}; long-time results (up to $t=70\,\mathrm{s}$) correspond to panels (c) and (d).}
\label{tab:observations}
\begin{tabular}{|l|l|l|}
\hline
\hspace{0.3cm}\text{($\gamma$, time)} & \hspace{2.5cm} \textbf{$B(t)/\sigma_0$} & \hspace{1.5cm}\textbf{$\mu(t)$} \\
\hline
{$\gamma=-1$, short} & Rapid initial squeezing 
& Quick rise toward $\pi/4$ \\
{$\gamma=-1$, long} & Oscillates with large spread and small squeezing 
&  Accumulates $\pi/2$ in each period\\
\hline
{$\gamma=0$, short} & Spreading 
& Gradual increase from $0$ to $\pi/4$ \\
{$\gamma=0$, long} & Instead of squeezing there is focusing effect  
& Accumulates $\pi/2$ \\
\hline
{$\gamma=1$, short} &  Rapid initial spreading 
& Slower growth  \\
{$\gamma=1$, long} & Oscillates with large spread and small squeezing
&  Accumulate $\pi/2$ in each period\\
\hline
\end{tabular}
\end{table}

 \subsection{Time and Correlation Effects for $\omega \gg \omega_0$}\label{sec:IIC}

We focus once more on the off-resonance dynamics, but now we consider the opposite end of the frequency spectrum. We examine the temporal evolution of the wavepacket in the regime where the natural frequency of the oscillator is much larger than the intrinsic frequency of the particle, i.e., \(\omega \gg \omega_0\), as well as the effects of the initial position-momentum correlation.

\begin{figure*}[!htb]
\centering
\includegraphics[scale=0.3]{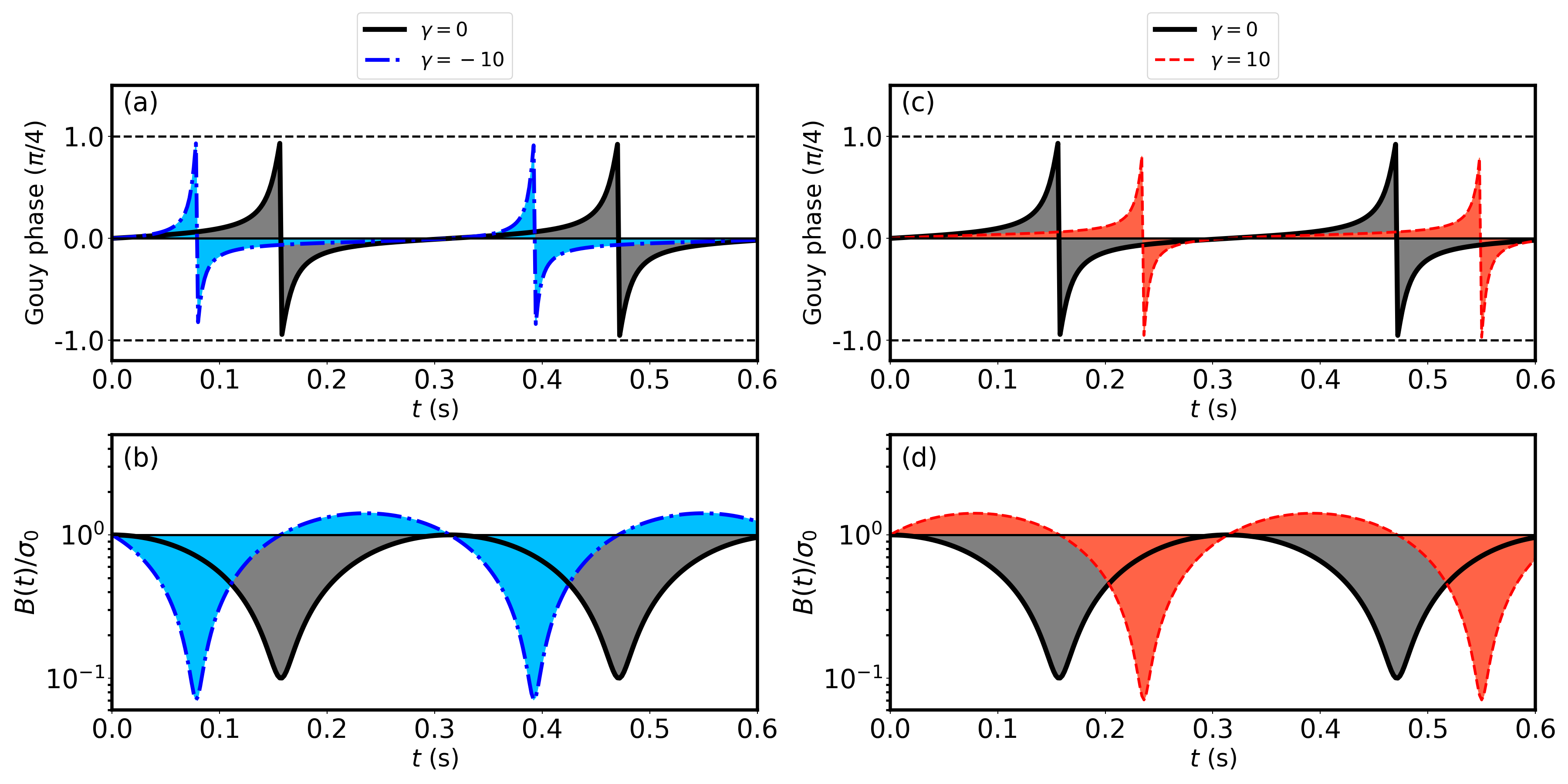}
\caption{Temporal behavior of the Gouy phase and wavepacket width when the oscillator's frequency is much larger than the intrinsic frequency of the particle  $\omega \gg \omega_0$. We consider  $\omega = 10$ \; \text{Hz} and $\omega_0 = 1 \; \text{Hz}$. Panels (a) and (b) compare the dynamics for $\gamma = -10$ with respect to $\gamma = 0$, while panels (c) and (d) illustrate the case of $\gamma = 10$ versus $\gamma = 0$. The width of the wavepacket exhibits temporal oscillations, with a total Gouy phase shift of $\pi/2\;\mathrm{rad}$ accumulated during the interval of time between two successive minima of the wavepacket width. The wavepacket width is always squeezed for $\gamma=0$. For $\gamma=10$ and $\gamma=-10$ the wavepacket squeezes and spread above its initial width. }\label{Figure3}
\end{figure*}

The Gouy phase (in units of \(\pi/4\;\mathrm{rad}\)) and the wavepacket width (in units of \(\sigma_0\)) as a function of time  in this regime are presented in Fig.~\ref{Figure3}. We consider \(\omega=10\;\mathrm{Hz}\), \(\omega_0=1\;\mathrm{Hz}\) and three different values of the correlation parameter \(\gamma\). In Fig.~\ref{Figure3}(a) and Fig.~\ref{Figure3}(b) we consider \(\gamma=0\) and \(\gamma=-10\). In Fig.~\ref{Figure3}(c) and Fig.~\ref{Figure3}(d) we consider \(\gamma=0\) and \(\gamma=10\). 

The width of the wavepacket oscillates in time, and a total Gouy phase shift of \(\pi/2\;\mathrm{rad}\) is accumulated between successive minima of the wavepacket width. Unlike the behavior observed in the regime where the oscillator's eigenfrequency is lower than the  intrinsic frequency of the particle, the width of the wavepacket remains squeezed for \(\gamma=0\), with its width always smaller than or equal to the initial width. For \(\gamma=10\) and \(\gamma=-10\) the wavepacket undergoes squeezing, although there exist time intervals where its width exceeds the initial value.  Here, the changes produced by \( |\gamma|=1 \) is very small, therefore we consider large values of \( |\gamma| \) in order to see the initial position-momentum correlations effect in this frequency regime.
For \(\gamma = 0\), the time at which the wavepacket reaches its maximum width coincides with the time at which the Gouy phase undergoes changes in both concavity and sign. 
Moreover, for \(\gamma\neq0\) the time when the wavepacket is maximum corresponds to the instant when the Gouy phase concavity changes. 

Interestingly, the total Gouy phase of \(\pi/2\;\mathrm{rad}\) is acquired in each period for any value of \(\gamma\) when the wavepacket evolves from a point of maximum squeezing (minimum width) until the next point of maximum squeezing. This way of accumulating the Gouy phase is different from the usual focusing process, where a light or matter-wave beam first converges to the focus and then spreads out \cite{Paz2}. Moreover, the rapid accumulation of the Gouy phase is advantageous for quantum metrology applications \cite{Gu}.

The approximated expressions for the wavepacket width and Gouy phase in the regime discussed above are given, respectively, in Eq.~(\ref{eq:B_high_frequency}) and Eq.~(\ref{eq:gouy_high_frequency}) of Appendix \ref{ap:Gouy_expressions}. We summarize the key behaviors of the wavepacket width and the Gouy phase under different correlation effects in the 
$\omega\gg\omega_0$ regime in Table III.

\begin{table}[h]
\centering
\caption{Key behaviors in the off-resonance regime ($\omega \gg \omega_0$).}

\begin{tabular}{|l| p{3.0cm} |p{3.8cm}| p{3.5cm}|}
\hline
\multicolumn{4}{|c|}{\textbf{Off-Resonance Regime: } $\omega \gg \omega_0$}\\
\hline
\textbf{$\gamma$} 
& \textbf{Gouy Phase} 
& \textbf{Wavepacket Width} 
& \textbf{Correlation Effects}\\
\hline
$0$ 
& $\pi/2$ per cycle 
& Remains squeezed 
& Minimum width reached at regular intervals \\
\hline
$+10$ 
& $\pi/2$ per cycle 
& Strong oscillations; width exceeds a little the initial value 
& Minimum width reached later than for $\gamma=0$ \\
\hline
$-10$ 
& $\pi/2$ per cycle 
& Strong oscillations; width exceeds a little the initial value 
& Minimum width reached earlier than for $\gamma=0$ \\
\hline
\end{tabular}
\end{table}

\pagebreak

\subsection{Time and Correlation Effects in the Resonance $\omega=\omega_0$}\label{sec:IID}

In this section, we study the dynamics of the Gouy phase and wavepacket width in the resonant regime, where the initial frequency of the particle matches the natural frequency of the confining harmonic oscillator.
In this scenario, the results are strongly influenced by the position-momentum correlation parameter \(\gamma\). It comes from the fact that, if there is no position-momentum correlation, the initial Gaussian wavepacket is an eigenstate of the harmonic oscillator and would acquire only a global phase.
Notably, by varying \(\gamma\), one can observe characteristics of a slowly diffracting wavepacket. Moreover, we note that quantum parametric resonance can also be investigated in other contexts involving time-dependent potentials, where it has been shown to play a key role in phenomena such as quantum amplification, squeezing, and stability analysis \cite{Stefan_Weigert_2002,FACCHI2001117}. Then, we consider \(\omega=\omega_0\), which produces for the wavepacket width and the Gouy phase, respectively, the following results
\begin{equation}
	B^2(t,\gamma)=\sigma_0^2\Big[\sin^2\omega t+(\gamma\sin\omega t+\cos\omega t)^2\Big],
\end{equation}  
and
\begin{equation}\label{eq:phase_ressonance}
	\mu(t,\gamma)=\frac{1}{2}\arctan\Bigg(\frac{\sin\omega t}{\gamma\sin\omega t+\cos\omega t}\Bigg).
\end{equation}
Note that for an initially uncorrelated Gaussian state \(\gamma=0\), the wavepacket width is time independent, i.e., \(B(t,\gamma=0)=\sigma_0\), and the Gouy phase is linear with the propagation time, i.e., \(\mu(t,\gamma=0)=\frac{\omega t}{2}\), in agreement with the Gouy phase of a Bessel beam \cite{R7}. Therefore, in resonance (\(\omega=\omega_0\)) and for \(\gamma=0\), the wavepacket is nondiffracting and its Gouy phase does not follow the conventional arctan profile characteristic of diffracting Gaussian wavepackets.

\begin{figure*}[!ht]
\centering
\includegraphics[scale=0.20]{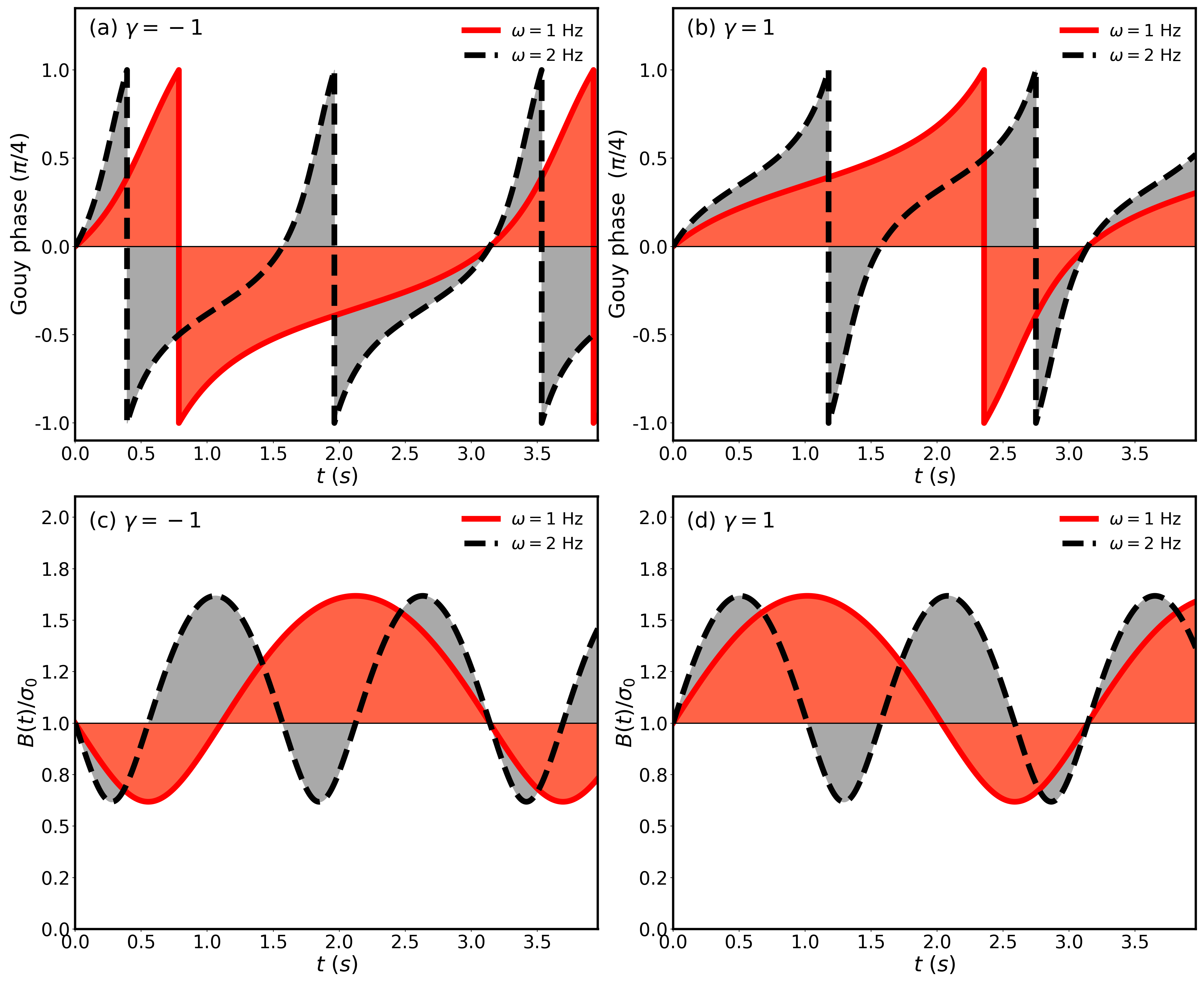}
\caption{Gouy phase and wavepacket width in the resonance as a function of propagation time $t$, for three distinct values of the initial correlation $\gamma$. Panels (a) and (c) correspond to $\gamma =-1$, and panels (b) and (d) to $\gamma =1$. It can be observed that as the frequency increases, the period of the Gouy phase decreases. Additionally, it is noteworthy that the period, given by $\pi/\omega$, remains independent of the initial correlation parameter $\gamma$, but it is affected as a time translation because of the changing in the squeezing direction. In the resonance the wavepacket spreads and squeezes, with spreading effect being more intense. The total Gouy phase of $\pi/2$ acquired in each period is a consequence of spreading and squeezing.}\label{Figure4}
\end{figure*}

In Fig.~\ref{Figure4}, we show the Gouy phase (in units of \(\pi/4\;\mathrm{rad}\)) and the wavepacket width (in units of \(\sigma_0\)) in the resonant regime, as a function of propagation time \(t\), for three distinct values of the initial correlation \(\gamma\). Fig.~\ref{Figure4}(a) and Fig.~\ref{Figure4}(c) correspond to \(\gamma=-1\), and Fig.~\ref{Figure4}(b) and Fig.~\ref{Figure4}(d) to \(\gamma=1\). It can be observed that as the frequency increases, the period of the Gouy phase decreases, i.e., the total Gouy phase of \(\pi/2\;\mathrm{rad}\) can be acquired in a small interval of time. 
Moreover, it is important to highlight that the period, given by \(\pi/\omega\), remains independent of the initial correlation parameter \(\gamma\), but it is affected by a time shift due to changes in the squeezing direction. Different from the regimes explored above, respectively, \(\omega\ll\omega_0\) and \(\omega\gg\omega_0\), in the resonant regime, although the spreading is more pronounced, the difference between spreading and squeezing is less extreme.

Again, the total Gouy phase of $\pi/2$ acquired in each period is a consequence of spreading and squeezing. 
The maxima values of the wavepacket width give the values of time for which the Gouy phase changes its concavity.
Inside each period, the total Gouy phase is acquired in two parts of the time evolution.

\begin{itemize}

\item For $\gamma=-1$, the Gouy phase of $\pi/4\;\mathrm{rad}$ is accumulated when the wavepacket evolves from a point of maximum squeezing (minimum width) until the second value of time for which the wavepacket width becomes equal to the width of the initial wavepacket. 
Another portion of $\pi/4\;\mathrm{rad}$ is acquired when the wavepacket squeezes until the next point of maximum squeezing.

\item For $\gamma=1$, $\pi/4\;\mathrm{rad}$ is obtained when the wavepacket evolves from a point of maximum squeezing until the first value of time for which the wavepacket width becomes equal to the initial width. The last portion of $\pi/4\;\mathrm{rad}$ is accumulated when the wavepacket evolves from this point to the next point of maximum squeezing.

\end{itemize}

The difference in the results of the Gouy phase produced by $\gamma=-1$ and $\gamma=1$ happens because the direction of squeezing is changed such that for $\gamma=-1$ the wavepacket starts the evolution squeezing and for $\gamma=1$ it starts spreading.

\begin{figure*}[!htb]
\centering
\includegraphics[scale = 0.40]{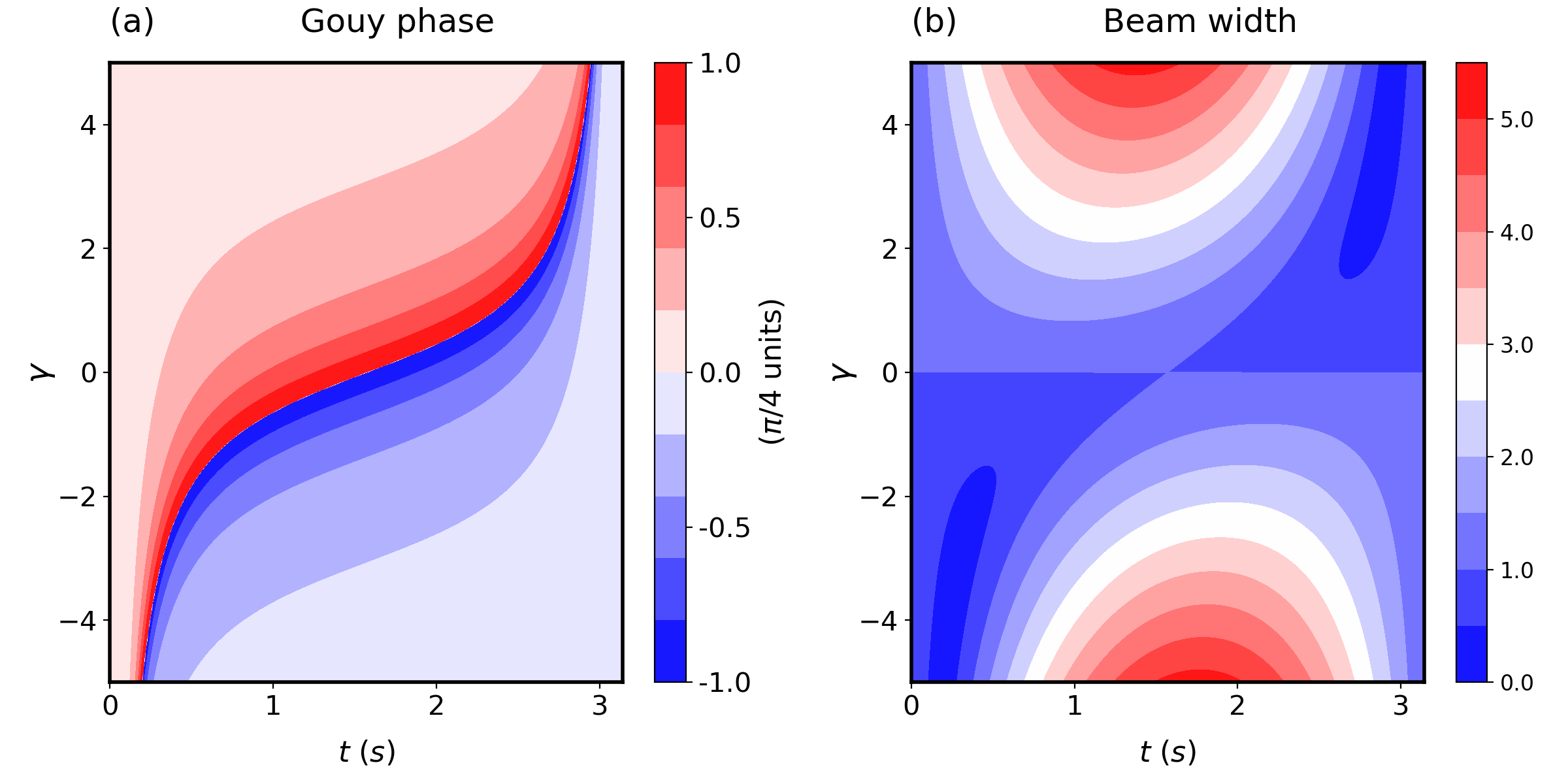}
\caption{One period of the (a) Gouy phase and (b) wavepacket width contour plots as a function of correlation parameter $\gamma$ and the propagation time $t$ for $\omega=\omega_0=1\;\mathrm{Hz}$. Gouy phase is expressed in units of $\pi/4\;\mathrm{rad}$ and the wavepacket width in units of $\sigma_0$. There are regions of squeezing of the wavepacket where $B(t,\gamma)<\sigma_0$ (more visible as dark blue color) and regions of spreading where $B(t,\gamma)>\sigma_0$ (more visible as dark red color). The Gouy phase accumulates the total quantity of $\pi/2\;\mathrm{rad}$ when the correlation parameter and the propagation time change.}\label{Figure5}
\end{figure*}

Next, to better understand the effects of time evolution and the initial position-momentum correlations on the acquired Gouy phase, we exhibit the contour plots of this phase in Fig.~\ref{Figure5}. In Fig.~\ref{Figure5}(a), we present the Gouy phase, while in Fig.~\ref{Figure5}(b), we display the contour plot of the wavepacket width as functions of the correlation parameter \(\gamma\) and the propagation time \(t\), for \(\omega=\omega_0=1\;\mathrm{Hz}\). As before, the Gouy phase is expressed in units of \(\pi/4\;\mathrm{rad}\) and the wavepacket width in units of \(\sigma_0\). We can observe by the color mapping that there are regions of squeezing of the wavepacket where \(B(t,\gamma)<\sigma_0\) (represented in dark blue) and regions of spreading where \(B(t,\gamma)>\sigma_0\) (depicted in dark red). The contour plot reveals that while squeezing occurs for both positive and negative \(\gamma\), the variation profile differs, reflecting the dependence of the squeezing direction on the sign of \(\gamma\). The Gouy phase accumulates the total quantity of \(\pi/2\;\mathrm{rad}\) when the correlation parameter and the propagation time change. The accumulated portion of \(-\pi/4\;\mathrm{rad}\) is related to the wavepacket squeezing and the accumulated portion of \(\pi/4\;\mathrm{rad}\) to the wavepacket spreading. Also, we can see that squeezing and spreading effects are intensified as \(|\gamma|\) increases. Only one period is shown for clarity. A panoramic overview of the correlation effects on the Gouy phase and wavepacket width in the resonance regime is shown in Table IV.

\begin{table}[h]
\centering
\caption{Key behaviors in the resonant regime ($\omega = \omega_0$). 
         The Gouy phase accumulates $\pi/2$ each period, and nonzero $\gamma$ shifts the timing and extent of squeezing/spreading.}
\begin{tabular}{|l| p{3.0cm} |p{3.8cm}| p{3.5cm}|}
\hline
\multicolumn{4}{|c|}{\textbf{Resonance Regime: } $\omega = \omega_0$}\\
\hline
\textbf{$\gamma$} 
& \textbf{Gouy Phase} 
& \textbf{Wavepacket Width} 
& \textbf{Correlation Effects}\\
\hline
$-1$ 
& $\pi/2$ per period 
& Initially narrower, then spreads and resqueezes 
& Negative correlation shifts squeezing earlier \\
\hline
$0$ 
& Linear profile
& No changing
& No effect \\
\hline
$+1$ 
& $\pi/2$ per period 
& Initially wider, then squeezes and re-expands 
& Positive correlation delays minimum width \\
\hline
\end{tabular}
\end{table}

\subsection{Resonant Dynamics in the Weak Position-Momentum Correlation Limit and Nondiffraction-Like Wavepackets}\label{sec:weak_correlation}

We explore further into the resonance regime,  focusing on scenarios where the wavepacket exhibits only minor temporal variations from its initial state.
It still accumulates the total Gouy phase of \(\pi/2\;\mathrm{rad}\), while retaining an approximately linear temporal profile. 
In fact, as pointed out in \cite{PabloVaveliuk}, a real beam whose Gouy phase is close to that linear evolution in a given range will have nondiffracting properties in such a range.  
Here, the initial position-momentum correlation represented by the parameter \(\gamma\) governs such behavior.  
In other words, the departure from the linear dependence on \(t\) of the Gouy phase will determine the degree of nondiffracting behavior of such a wavepacket.  
In the context of light waves, the Gouy phase of nondiffracting beams has been studied theoretically and experimentally \cite{R7,R8}.  
Such light beams offer applications in biomedicine, microscopy, manipulation of microscopic particles, particle manipulation, precision measurements, and enhanced image resolution and contrast \cite{remosense,bio,R9,R10,R12}.  
Thus, given the importance and the vast range of applications of nondiffracting (or almost nondiffracting) beams in the field of light, it is crucial to rigorously determine the propagation regimes and conditions under which a correlated Gaussian matter-wavepacket, with finite energy and spatial extent, exhibits nondiffracting behavior.

\begin{figure*}[!htb]
\centering
\includegraphics[scale=0.37]{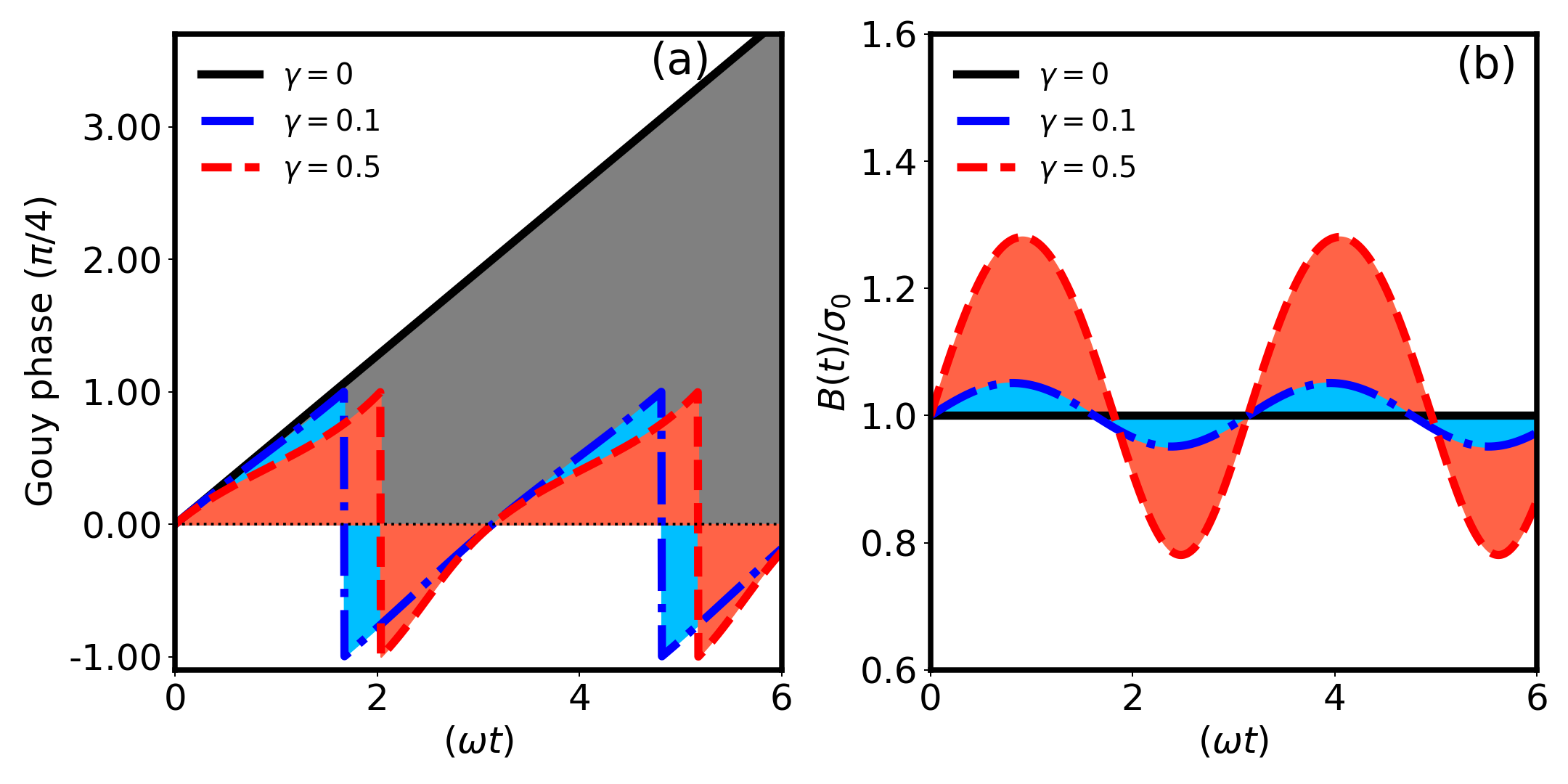}
\caption{Gouy phase and wavepacket width as functions of $\omega t$ for small values of correlation parameter $\gamma$. Panel (a) is the Gouy phase and panel (b) is the wavepacket width. The black solid line corresponds to $\gamma=0$, the blue dash-dot line to $\gamma=0.1$ and the red dashed red line to $\gamma=0.5$.}\label{Figure6}
\end{figure*}

The regime of nearly nondiffracting wavepackets can be achieved in the resonance condition (\(\omega=\omega_0\)) and for \(\gamma\ll1\), where the Gaussian wavepacket width will undergo only slight oscillations about its initial width.
In this limit, we obtain the following expressions for the Gouy phase and the wavepacket width
\begin{equation}
	\mu (t) \approx \frac{\omega t}{2} - \frac{1}{2} \sin^2 (\omega t) \;\gamma + \mathcal{O}\! \left(\gamma^{2}\right),
\end{equation}
and
\begin{equation}
	B(t) \approx \sigma_0 + \frac{1}{2}\sigma_0\sin (2\omega t) \;\gamma + \mathcal{O}\! \left(\gamma^{2}\right).
\end{equation}
We display their behavior as functions of \(\omega t\) for small values of the correlation parameter \(\gamma\) in Fig.~\ref{Figure6}. In Fig.~\ref{Figure6}(a) we show the Gouy phase and in Fig.~\ref{Figure6}(b) the wavepacket width. The black solid line corresponds to \(\gamma=0\), the blue dash-dot line to \(\gamma=0.1\) and the red dashed line to \(\gamma=0.5\). For \(\gamma=0\) the wavepacket is nondiffracting, and the corresponding Gouy phase increases linearly with time. For small values of \(\gamma\), the wavepacket width presents small oscillations around the initial width while the phase deviates slightly from the linear profile but preserves the key features of the Gouy phase typical of diffracting Gaussian wavepackets, with a total phase shift of \(\pi/2\;\mathrm{rad}\) in each period. For even smaller values of \(\gamma\), such deviation of the wavepacket width and Gouy phase is further reduced, suggesting that an almost nondiffracting wavepacket can be engineered.  
This kind of wavepacket can be useful in applications similar to those of nondiffracting light beams, such as nearly nondiffracting neutron or electron beams. Also, such wavepackets can be useful to reduce the effects of decoherence associated with wavepacket spreading \cite{decoherence}.

\section{Application: Quantum Metrological Role of the Gouy Phase and Squeezing Dynamics}\label{sec: Fisher}

 While squeezing and the Gouy phase are individually well-established resources in quantum technologies, to the best of our knowledge, their explicit interconnection, and in particular the ability to tune squeezing via controlled Gouy-phase evolution, has not previously been explored in the context of matter waves.
 In this section, we illustrate this novelty in the concrete setting of quantum metrology by evaluating the Classical and Quantum Fisher Information to quantify precision limits in parameter estimation. This approach highlights how tuning the Gouy phase, through its link to wavepacket squeezing, can directly enhance the information content of quantum states, providing a clear example of the technological potential of Gouy-phase–based control and opening new avenues for phase-sensitive quantum control strategies.

 We present a concise overview of quantum metrology, with particular emphasis on the roles of Classical Fisher Information (CFI) and Quantum Fisher Information (QFI) in the context of parameter estimation within our model.
For simplicity, in what follows we focus on analyzing the precision limits, that is, the bounds on the achievable accuracy for estimating the oscillator frequency in the resonant case $\omega = \omega_0$. For the evolved quantum state given in Eq.~(\ref{eq:evolved_state}), the position probability distribution takes the form
$P(x \mid \omega) = |\psi(x, t)|^2 = \frac{1}{B(t, \omega) \sqrt{\pi}} \exp\left[-\frac{x^2}{B(t, \omega)^2}\right]$,
and the corresponding evolved, dimensionless covariance matrix is given by:
\begin{equation}\label{eq:evolved_cov_matrix}
    \boldsymbol{\sigma}(t) =\begin{pmatrix} \;
\sigma_{xx}(t) & \sigma_{xp}(t)\\
\sigma_{xp}(t) & \sigma_{pp}(t) \;
\end{pmatrix}, 
\end{equation}
where

\begin{gather}
    \sigma_{xx}(t) = \frac{1}{2} \left[ \sin^2\omega t + \left( \gamma \sin\omega t + \cos\omega t \right)^2 \right]; \;\;\; \sigma_{xp}(t) = \sigma_{px}(t) =  \frac{1}{2} \gamma \left[ \gamma \sin\omega t \cos\omega t + \cos2\omega t \right]; \nonumber \\
   \sigma_{pp}(t) =  \frac{1}{2 \left[ \sin^2\omega t + \left( \gamma \sin\omega t + \cos\omega t \right)^2 \right]} 
\left\{ 1 +  \gamma^2 \left[ \gamma \sin\omega t \cos\omega t + \cos2\omega t \right]^2 \right\}.  
 \end{gather}
From these quantities, we can compute the Classical Fisher Information (QFI) $\mathcal{F}_{\omega}^{\;\text{C}}$ for position measurements to estimate the frequency $\omega$, using the following expressions (see Appendix~\ref{Ap:Fisher} for more details):
\begin{gather}
\mathcal{F}_{\omega}^{\;\text{C}} = \int_{-\infty}^{\infty} dx\, P(x|\omega) \left( \frac{\partial \ln P(x|\omega)}{\partial \omega} \right)^2
=  \int_{-\infty}^{\infty} dx\, \frac{1}{P(x|\omega)} \left( \frac{\partial P(x|\omega)}{\partial \omega} \right)^2 \nonumber\\
\mathcal{F}_{\omega}^{\;\text{C}} = 
\frac{t^2 \gamma^2 \left(2 \cos(2 t \omega) + \gamma \sin(2 t \omega)\right)^2}{8 \sqrt{2} \left[ \sin^2(t \omega) + \left(\cos(t \omega) + \gamma \sin(t \omega)\right)^2 \right]^{9/2} \left( \frac{1}{2 + \gamma^2 - \gamma^2 \cos(2 t \omega) + 2 \gamma \sin(2 t \omega)} \right)^{5/2}}, \label{eq:CFI}
\end{gather}
\begin{figure*}[!ht]
\centering
\includegraphics[scale=0.37]{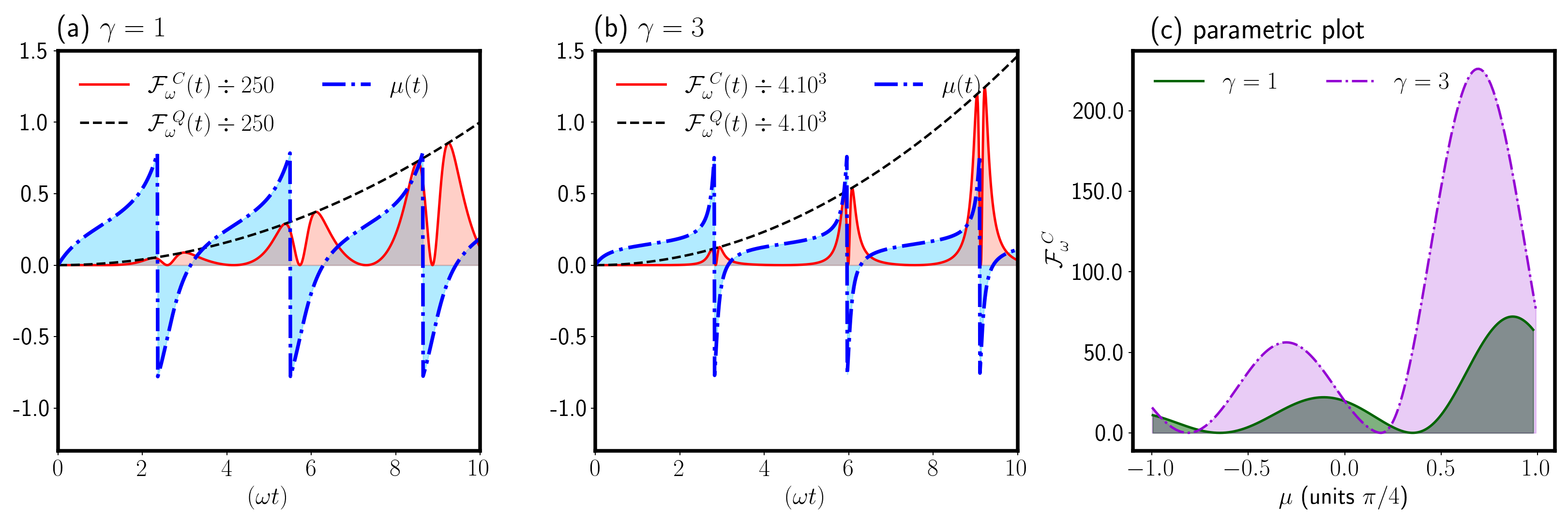}
\caption{Classical Fisher information (CFI) $\mathcal{F}{\omega}^{\;\text{C}}$, Quantum Fisher information (QFI) $\mathcal{F}{\omega}^{\;\text{Q}}$, and Gouy phase $\mu$ as functions of the dimensionless time $\omega t$. Panel (a) shows the results for an initial position–momentum correlation parameter $\gamma = 1$, while panel (b) corresponds to $\gamma = 3$. The CFI attains the maximum value of the QFI near the time at which the Gouy phase $\mu$ changes sign, a behavior associated with the squeezing of the wavepacket width, which enhances the information content. Notably, the greater the temporal variation of the Gouy phase, the more pronounced the gain in information. Additionally, stronger initial position–momentum correlations lead to higher values of the corresponding Fisher information, as evidenced by the comparison between panels (a) and (b). Panel (c) displays a parametric plot of $\mathcal{F}_{\omega}^{\;\text{C}}$ over one period of the Gouy phase $\mu$, illustrating that the maximum information can be achieved through a suitable choice of Gouy phase and correlation $\gamma$. For visualization purposes, the curve for $\gamma = 3$ has been rescaled by a factor of $1/9$ in panel (c). }\label{Figurefisher_mu}
\end{figure*}

\noindent while the corresponding Quantum Fisher Information (QFI) $\mathcal{F}_{\omega}^{\;\text{Q}}$ is given by:
\begin{equation}
\mathcal{F}_{\omega}^{\;\text{Q}} = \frac{\text{Tr}\{[\boldsymbol{\sigma}^{-1}(t)\partial_{\omega}\boldsymbol{\sigma}(t)]^{2}\}}{4}
\nonumber \\
= \frac{1}{2} t^2 \gamma^2 \left(4 + \gamma^2\right). \label{eq:QFI}
\end{equation}
Here, we have set \( \mu = 1 \) since our state is pure, taken the displacement vector as \( \boldsymbol{d} = \boldsymbol{0} \), and consequently suppressed the second term in Equation~(\ref{QFI_gaussian}), as discussed in \cite{serafini2017quantum} for pure states.
 The QFI provides the ultimate bound on the maximum precision with which the parameter \( \omega \) can be estimated, as constrained by the principles of quantum mechanics \cite{Helstrom1969,Holevo_book,Braunstein_CavesPRL1994,Milburn1996}. It also serves as an upper bound to the Classical Fisher Information, that is, $ \mathcal{F}_{\omega}^{\;\text{C}} \leq \mathcal{F}_{\omega}^{\;\text{Q}}$.

Figure~\ref{Figurefisher_mu} illustrates the behavior of the classical Fisher information (CFI) $\mathcal{F}{\omega}^{\;\text{C}}$, quantum Fisher information (QFI) $\mathcal{F}{\omega}^{\;\text{Q}}$, and the Gouy phase $\mu$ as functions of the dimensionless time $\omega t$ for two different values of the initial position–momentum correlation parameter $\gamma$. For $\gamma = 1$ [Fig. \ref{Figurefisher_mu}(a)], the CFI reaches the maximum value allowed by the QFI near the point where the Gouy phase changes sign, indicating a strong connection between phase dynamics and information gain. When the initial correlation is increased to $\gamma = 3$ [Fig. \ref{Figurefisher_mu}(b)], both the CFI and QFI attain higher values, demonstrating that stronger initial correlations enhance the sensitivity of the system to changes in $\omega$. This enhancement is attributed to the squeezing of the wavepacket width, which effectively concentrates the probability distribution and increases the information content. Moreover, the temporal rate of change of the Gouy phase is directly linked to the gain in information, the faster the variation, the greater the CFI. This behavior is further explored in Fig.~\ref{Figurefisher_mu}(c), which presents a parametric plot of $\mathcal{F}_{\omega}^{\;\text{C}}$ as a function of the Gouy phase over one oscillation period. The results show that the maximum classical Fisher information can be attained through an appropriate choice of both the Gouy phase and the initial correlation parameter $\gamma$. For clarity in visualization, the curve corresponding to $\gamma = 3$ has been rescaled by a factor of $1/9$.

The results in Fig.~\ref{Figurefisher_mu} show that the temporal behavior of the Gouy phase is directly linked to the attainable precision in estimating system parameters, as quantified by the CFI and QFI. Maxima of the CFI occur near points where the Gouy phase changes sign, coinciding with enhanced squeezing of the wavepacket width. Stronger initial position–momentum correlations amplify both the Gouy phase variation and the Fisher information, demonstrating that the phase–squeezing interplay can be harnessed to optimize sensitivity. Our findings establish a direct conceptual link between Gouy-phase–induced squeezing and optimal parameter estimation, opening a pathway to phase-sensitive quantum control protocols where the Gouy phase is treated as an actively engineered resource. In this sense, our framework provides a theoretical baseline for developing a potential new tool for quantum technologies, with immediate implications for quantum metrology and possible extensions to open or driven systems.

\section{CONCLUSION}\label{sec:conclusion}

We studied the quantum dynamics of a Gaussian wavepacket with position-momentum correlations, characterized by the correlation parameter \(\gamma\), for a particle of intrinsic frequency $\omega_0$ confined in a harmonic potential with frequency \(\omega\), focusing on the behavior of the Gouy phase and wavepacket width. The case \(\gamma=0\) corresponds to the fundamental state of a harmonic oscillator of frequency \(\omega_0\). These three parameters govern the wavepacket dynamics and  lead to interesting effects such as squeezing.

We discussed the wavepacket and Gouy phase profiles for different regimes of confinement. In the off-resonance regimes \(\omega \ll \omega_0\) and \(\omega \gg \omega_0\), we observed that the wavepacket exhibits distinct squeezing and spreading behaviors, which are strongly modulated by the position-momentum correlations. For \(\omega \ll \omega_0\), the wavepacket predominantly spreads, with squeezing effects becoming more pronounced for larger values of \(\gamma\) and no squeeze for \(\gamma=0\). Although the squeezing effect is small when the frequency of the harmonic oscillator is much smaller than the intrinsic frequency of the particle, it is still crucial for the acquired Gouy phase. 
Conversely, in the regime where the oscillator's frequency is much larger than the intrinsic frequency of the particle \(\omega \gg \omega_0\), the wavepacket remains squeezed for \(\gamma = 0\), while for \(\gamma \neq 0\), it alternates between squeezing and spreading. Notably, the Gouy phase accumulation is closely tied to the wavepacket's dynamics, with squeezing and spreading effects playing complementary roles in its evolution. These behaviors directly influence the accumulation of the Gouy phase, which reaches a total of \(\pi/2\;\mathrm{rad}\) over specific intervals of propagation time. In the resonant regime \(\omega = \omega_0\), the system exhibits unique characteristics, particularly for small values of \(\gamma\). Here, the wavepacket width remains nearly constant, and the Gouy phase evolves linearly with time, mimicking the behavior of nondiffracting beams. For \(\gamma \ll 1\), the wavepacket undergoes minimal spreading and squeezing, suggesting the potential of engineering almost nondiffracting matter-wavepackets. Thus, we showed that one can design a wavepacket robust against decoherence with significant potential for precision measurements, microscopy, and the preservation of quantum coherence.
A significant difference was observed compared to the free evolution of matter waves in one dimension, where the total acquired Gouy phase is only \(\pi/4\;\mathrm{rad}\). In the present study, the matter waves evolving in one dimension acquired the total Gouy phase of \(\pi/2\;\mathrm{rad}\) in each interval of time, corresponding to the evolution from the minimum width (maximum squeezing) back to the same state. 

Our analysis of the Classical and Quantum Fisher Information reveals that the temporal evolution of the Gouy phase through its intimate link to wavepacket squeezing, can directly influence the precision limits of parameter estimation. Peaks in the Fisher information occur near points where the Gouy phase changes sign, a feature tied to enhanced squeezing and stronger position-momentum correlations. This establishes a concrete connection between Gouy-phase–induced squeezing and quantum metrology, demonstrating that the Gouy phase can be treated not merely as a passive geometric quantity, but as an actively tunable resource. While quantum metrology provides a clear example, the underlying mechanism has broader implications for quantum technologies, where engineered Gouy-phase control could form the basis for new phase-sensitive quantum protocols.  In this sense, our framework serves as a theoretical foundation for developing a potential new tool for quantum technology.

In this work, we have focused on an idealized, closed-system model, enabling an analytically tractable investigation of the Gouy phase and its connection to squeezing. In a realistic scenario, however, continuous couplings to stray optical and phonon modes, electromagnetic noise, or unintended time-dependent trap variations can induce gradual damping and loss of phase coherence, while discrete events such as residual-gas collisions disrupt correlations and broaden the wavepacket.
Losses have not been included in the present analysis, as our primary objective is to first elucidate the Gouy-phase–squeezing connection in its most fundamental form, thereby establishing a foundation for future extensions of the model to open quantum systems and explicitly driven traps.

The Gouy phase, commonly observed when waves undergoes spatial confinement, here is manifested, in part, as a consequence of the squeezing of matter waves. Within this framework, the Gouy phase and the wavepacket width are linked through Eq. (\ref{eq:GouyRate}), allowing the Gouy phase to be indirectly determined through precise measurements of the motional wavepacket's spatial width. Such position measurements, accessible in ion-trap experiments, may thus provide an experimental route to probing the temporal evolution of the Gouy phase.
Given the broad range of applications for squeezed states, controlling how squeezing shapes the Gouy phase, and \textit{vice-versa}, offers a direct way to tune the phase evolution of confined matter waves. In our framework, this means that by adjusting the initial position–momentum correlations and the trap parameters, one could design states whose Gouy phase accumulates at a prescribed rate while maintaining minimal diffraction.  Looking forward, these concepts could be extended to more complex or time-dependent potentials, enabling the exploration of driven squeezing–phase dynamics, as well as to experimental realizations in ultracold atomic setups or tailored optical analogues, where direct measurement of the wavepacket width could serve as a proxy for the Gouy phase.

\begin{acknowledgments}

T.M.S. O Thanks CAPES (Brazil) for the financial support. L.S.M. acknowledges the Federal University of Piau\' i for providing the workspace. F. C. V. de Brito acknowledges the financial support from Horizon Europe, the European Union's Framework Programme for Research and Innovation, SEQUOIA project, under Grant Agreement No. 101070062. M.S. acknowledges a research grant 302790/2020-9 from CNPq.  I.G.P. acknowledges Grant No. 306528/2023-1 from CNPq. 

\end{acknowledgments}

\section*{Ethics declarations}
\textbf{Competing interests}. The authors declare no competing interests.

\appendix

\section{Relation between Position-Momentum Correlation and Squeezing}\label{Wigner}

Consider the evolution of a position-momentum correlated Gaussian wavepacket for a massive particle in a harmonic potential. The initial state described by the correlated Gaussian wavefunction is given by \cite{DODONOV1980PLA}
\begin{equation}
	\psi_0(x') = \frac{1}{\sqrt{\sigma_0 \sqrt{\pi}}} \exp\left[-\frac{x'^2}{2\sigma_0^2} + \frac{i\gamma\, x'^2}{2\sigma_0^2}\right],
	\label{initial_packet}
\end{equation}
where $\omega_0$ is the intrinsic spread frequency of the initial wavepacket. This state represents an approximately localized state in another potential, which does not necessarily correspond to an eigenstate of the harmonic oscillator, within which we will consider the particle to be confined. For the initial quantum state under consideration, the uncertainties in position and momentum are given by $\sigma_{xx}(0)= \langle \psi_0 |\hat{x}^2| \psi_0 \rangle - \langle \psi_0 |\hat{x}| \psi_0 \rangle=\sigma_{0}^2/2$ and $\sigma_{pp}(0)=\langle \psi_0 |\hat{p}^2| \psi_0 \rangle - \langle \psi_0 |\hat{p} \psi_0 \rangle=(1+\gamma^{2})\hbar^2/2\sigma_{0}^2$, respectively. The associated covariance between position and momentum is defined as $\sigma_{xp}(0)= \langle \psi_0 |( \hat{x}\hat{p} +\hat{p}\hat{x})/2 | \psi_0 \rangle -\langle \psi_0 |\hat{x} |\psi_0 \rangle \langle \psi_0 |\hat{p} |\psi_0 \rangle =\hbar\gamma/2.$
To quantify the degree of statistical correlation between the position operator $\hat{x}$ and the momentum operator $\hat{p}$, the dimensionless Pearson correlation coefficient is introduced as $P=\sigma_{xp}/\sqrt{\sigma_{xx}\sigma_{pp}}$, constrained within the interval $-1\leq P\leq1$. Within this framework, the parameter $\gamma$ can be expressed as a function of $P$ by $\gamma=P/\sqrt{1-P^{2}}$, with $\gamma$ ranging over the entire real line $(-\infty\leq \gamma\leq\infty)$. This expression highlights the physical interpretation of $\gamma$ as a parameter that encodes the initial correlations between the conjugate observables $\hat{x}$ and $\hat{p}$. In the particular case of $\gamma=0$, the system reduces to an uncorrelated initial Gaussian wavepacket. The study of such position-momentum correlations was originally conducted in Ref.~\cite{bohm1951quantum}.

Notably, in defining the covariance and correlation coefficient, the analysis explicitly incorporates the quantum mechanical nature of the operators $\hat{x}$ and $\hat{p}$, which satisfy the canonical commutation relation $[\hat{x},\hat{p}]=i\hbar$. Furthermore, to ensure the hermiticity of the product operator, the symmetrization procedure is applied, resulting in the operator $(\hat{x}\hat{p} +\hat{p}\hat{x})/2$. From a practical perspective, the physical origin of the parameter $\gamma$ may be interpreted as arising from the propagation of an atomic beam through a transverse harmonic potential, which effectively acts as a thin lens, thereby imprinting a quadratic phase onto the initial state \cite{JanickeJMO1995}. Here, it influences the confined dynamics, leading to distinct modifications in the wavepacket evolution. Based on the preceding calculations of the uncertainties and the covariance between the position and momentum operators, the corresponding dimensionless covariance matrix for this initial correlated Gaussian state can be expressed as follows \cite{RevModPhys2012Lloyd}:
\begin{equation}\label{eq:iniial_cov_matrix}
    \boldsymbol{\sigma}(0) =\begin{pmatrix} \;
\frac{\sigma_{xx}}{\sigma_0^2} & \frac{\sigma_{xp}}{\hbar}\\
\frac{\sigma_{xp}}{\hbar} & \frac{\sigma_{pp}}{(\hbar^2/\sigma_0^2)} \;
\end{pmatrix}= \frac{1}{2}\begin{pmatrix} \;
1 & \gamma \\
\gamma & 1+\gamma^2 \;
\end{pmatrix}.
\end{equation}

A well-established relationship exists between position-momentum correlations and the phenomenon of squeezing~\cite{MarinhoPRA2020,Porto2024}. To elucidate this connection more clearly, consider the single-mode squeezed vacuum state defined as $|S(\zeta)\rangle = S(\zeta)|0\rangle$, where the covariance matrix is given by~\cite{LandiPRA2024}:
\begin{equation}\label{eq:squeezed_state}
    \boldsymbol{\sigma_S} = \frac{1}{2}\begin{pmatrix} \;
\cosh 2r - \sinh 2r \cos\phi & - \sinh 2r \sin\phi  \\
- \sinh 2r \sin\phi  & \cosh 2r + \sinh 2r \cos\phi \;
\end{pmatrix}.
\end{equation}
Here, $S(\zeta)$ denotes the squeezing operator with a complex squeezing parameter $\zeta = re^{i\phi}$, where $r$ characterizes the degree of squeezing and $\phi$ defines the orientation of the squeezing in phase space. By comparing the off-diagonal elements of the covariance matrix in Eq.\eqref{eq:squeezed_state} with $\sigma_{xp}(0)/\hbar=\gamma/2$ of the initial state \eqref{initial_packet}, one finds that the correlation parameter is explicitly given by $\gamma = -\sinh(2r)\sin\phi$. Consequently, when $\phi = 0$, we obtain $\gamma = 0$, implying that a nonzero position-momentum correlation arises only when the squeezed state is rotated in phase space.

On the other hand, it is important to emphasize that such correlations can also emerge dynamically, even in the absence of squeezing. For instance, a coherent state that undergoes free evolution governed by the Hamiltonian $\hat{H} = \hat{p}^2/2m$ acquires position-momentum correlations as a result of the system's temporal evolution rather than through initial state preparation. In this scenario, the covariance matrix evolves into the form~\cite{bohm1951quantum}:
\begin{equation}
    \boldsymbol{\sigma} = \frac{1}{2}\begin{pmatrix} \;
1+ \frac{t^2}{\tau^2}\;\;\; & \frac{t}{\tau}  \\
\frac{t}{\tau}  \;\;\;& 1\;
\end{pmatrix},\label{xp_dinamic}
\end{equation}
where $\tau$ represents a characteristic time scale associated with the evolution. The emergence of such correlations leads to a phase-space distribution that is predominantly confined within an elliptical region, as originally discussed in~\cite{bohm1951quantum}. Physically, this reflects the tendency for large position values to become correlated with large momenta over time, producing a stretched anisotropic phase-space profile aligned along a particular direction.

To elucidate the role of position-momentum correlations and their intrinsic connection to squeezing, we present phase-space representations of the quantum state via the Wigner function \cite{Wigner1932}, which provides a quasi-probability distribution in classical phase space $(x, p)$ and serves as a valuable tool for visualizing quantum states \cite{BanaszekPRA1998}. For pure states it is defined by \cite{leonhardt1997measuring}:  
\begin{equation}
    W(x,p) = \frac{1}{\pi\hbar}\int_{-\infty}^{\infty} dy \; e^{2ipy/\hbar} \psi^* (x+y) \psi (x-y).
\end{equation}
For any Gaussian state, the Wigner function assumes a Gaussian form and can be expressed explicitly as \cite{RevModPhys2012Lloyd}:
\begin{equation}
    W (\boldsymbol{r}) =\frac{ \exp\left[-\frac{1}{2}(\boldsymbol{r}-\boldsymbol{d})^{\text{T}}\boldsymbol{\sigma}^{-1}(\boldsymbol{r}-\boldsymbol{d})\right]}{(2\pi) \sqrt{\text{det}\boldsymbol{\sigma}}},
\end{equation}

\begin{figure*}[h]
\centering
\includegraphics[scale=0.4]{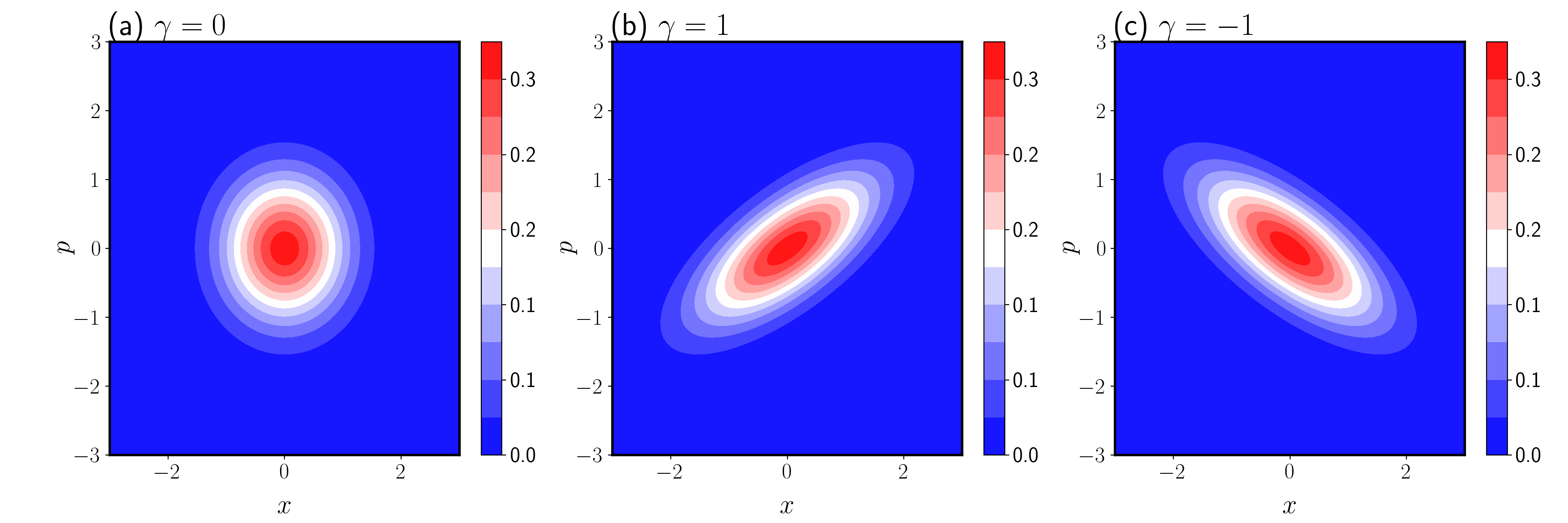}
\caption{Wigner function of the initial state for different values of the correlation parameter $\gamma$. For $\gamma=0$, the distribution is symmetric, while $\gamma =\pm 1 $ produce elliptical shapes rotated clockwise ($\gamma = 1$) or counterclockwise ($\gamma =-1$), showing the effect of position-momentum correlations on phase space orientation.  }\label{FigureWigner}
\end{figure*}

Where $\boldsymbol{r} = (x, p)$ denotes the vector of phase-space coordinates. Consequently, the quantum state can be uniquely characterized by its first- and second-order moments: the displacement vector $\boldsymbol{d}$ and the covariance matrix $\boldsymbol{\sigma}$, respectively, where
\begin{equation}
\boldsymbol{d} = ( \langle \hat{x} \rangle , \langle \hat{p} \rangle ),
\end{equation}
which, for our initial state (\ref{initial_packet}), is equal to zero. The figure \ref{FigureWigner} shows the Wigner function of the initial state for different values of the correlation parameter $\gamma$, illustrating how position-momentum correlations shape the phase space distribution. For $\gamma =0$, the Wigner function is symmetric with respect to the position and momentum axes, reflecting a minimum uncertainty Gaussian state with no correlation between the variables. When $\gamma = 1$, the Wigner function becomes an elliptic distribution rotated in the clockwise direction, indicating positive correlation between position and momentum. Conversely, for $\gamma = -1$, the ellipse is rotated counterclockwise, revealing a negative correlation. These rotations highlight how the phase space orientation encodes the nature of the correlations in the quantum state.

\section{ Expressions for the Gouy phase and wavepacket width far from the resonance}\label{ap:Gouy_expressions}

In the regime of low harmonic oscillator's frequency $\omega \ll \omega_0$ the expressions (\ref{eq:beam_width_nonressonant}) and (\ref{eq:gouy_nonressonant}) for the wavepacket width and Gouy phase can be approximated to
\begin{gather}
    B(t) \approx \sigma_0 \sqrt{1+2\gamma\omega_0 t+(1+\gamma^2)\omega_0^2 t^2 } -\frac{\sigma_0\Big[3+4\gamma\omega_0t+(1+\gamma^2)\omega_0^2t^2\Big]t^2}{6\sqrt{1+2\gamma\omega_0t+(1+\gamma^2)\omega_0^2t^2}} \; \omega^2 + \mathcal{O}\! \left(\omega^{4}\right),  \label{eq:B_low_frequency}
\end{gather}

and

\begin{gather}
        \mu (t) \approx \frac{1}{2}\arctan \Bigg(\frac{\omega_0 t}{1+\gamma\omega_0 t} \Bigg)  + \frac{\omega_0 t^3}{6+12\gamma\omega_0t+6\omega_0^2t^2(1+\gamma^2)}\; \omega^2 + \mathcal{O}\! \left(\omega^{4}\right). \label{eq:gouy_low_frequency}
\end{gather}
The first term of both expressions reproduces the free dynamics studied previously in \cite{lustosa2020irrealism}, defining \(\tau_0=\omega_{0}^{-1}\) as the temporal analogous of the Rayleigh length.

In the regime where the natural frequency of the confining oscillator is much larger than the intrinsic frequency of the particle  $\omega \gg \omega_0$ the expressions (6) and (8) for the wavepacket width and Gouy phase can be approximated to
\begin{equation}
    B(t) \approx \sigma_0|\cos(\omega t)|+ \frac{\sigma_0 \gamma\omega_0 |\cos(\omega t)| \sin(\omega t)}{\cos(\omega t)} \frac{1}{\omega}  + \mathcal{O}\! \left(\frac{1}{\omega^2}\right), \label{eq:B_high_frequency}
\end{equation}
and
\begin{equation}
 \mu(t) \approx\frac{\tan(\omega t)}{2}\left(\frac{\omega_{0}}{\omega}\right)-\gamma\frac{\tan(\omega t)^{2}}{2}\left(\frac{\omega_{0}}{\omega}\right)^{2} +\frac{(3\gamma^{2}-1)\tan(\omega t)^{3}}{6}\left(\frac{\omega_{0}}{\omega}\right)^{3}+\frac{(-\gamma^{3}+\gamma)\tan(\omega t)^{4}}{2}\left(\frac{\omega_{0}}{\omega}\right)^{4}+\mathcal{O}\! \left(\frac{\omega_{0}}{\omega}\right)^{5}. \label{eq:gouy_high_frequency}
\end{equation}.

\section{Fisher Information}\label{Ap:Fisher}

The Fisher information quantifies how sensitively a probability distribution responds to changes in an unknown parameter $\theta$ \cite{Fisher_1925}. For a set of outcomes $x_i$ distributed according to $P(x_i|\theta)$, the Classical Fisher Information (CFI) is defined as:
\begin{gather}
\mathcal{F}_{\theta}^{\;\text{C}} = \sum_{i} P(x_i|\theta)\left(\frac{\partial \ln[P(x_i|\theta)]}{\partial \theta}\right)^2
= \sum_{i} \frac{1}{P(x_i|\theta)} \left( \frac{\partial P(x_i|\theta)}{\partial \theta} \right)^2.
\end{gather}

A peaked probability distribution with respect to $\theta$ indicates high information content, while a flat one corresponds to poor sensitivity. The achievable precision is bounded by the Cramér-Rao inequality \cite{cramer1946}:
\begin{equation}
\Delta \theta \geq \frac{1}{\sqrt{n \mathcal{F}_{\theta}^{\;\text{C}}}},
\end{equation}
where $n$ is the number of experimental repetitions and $\Delta \theta$ is the standard deviation of the estimator. This bound holds for unbiased estimators \cite{Fisher_1925}. In the quantum framework, measurements are described using Positive Operator-Valued Measures (POVMs) ${\hat{A}(\kappa)}$, satisfying $\sum_{\kappa} \hat{A}(\kappa) = \hat{I}$. The outcome probabilities are given by:
$P(\kappa|\theta) = \text{Tr}[\hat{\rho} \hat{A}(\kappa)]$, which allows the CFI to be expressed in the quantum setting as:
\begin{equation}
\mathcal{F}_{\theta}^{\;\text{C}}(\hat{A}) = \sum_{\kappa} \frac{1}{P(\kappa|\theta)} \left( \frac{\partial P(\kappa|\theta)}{\partial \theta} \right)^2.
\end{equation}

Maximizing the CFI over all possible POVMs gives the Quantum Fisher Information (QFI) \cite{Helstrom1969,Holevo_book,Braunstein_CavesPRL1994,Milburn1996}:
\begin{equation}
\mathcal{F}_{\theta}^{\;\text{Q}} = \max_{\hat{A}} \mathcal{F}_{\theta}^{\;\text{C}}(\hat{A}),
\end{equation}
which sets the ultimate precision limit for estimating $\theta$. The inequality $\mathcal{F}_{\theta}^{\;\text{Q}} \geq \mathcal{F}_{\theta}^{\;\text{C}}$ always holds.

For single-mode Gaussian states, characterized by a covariance matrix $\boldsymbol{\sigma}$ and displacement vector $\boldsymbol{d}$, the QFI can be computed analytically as shown in Refs.~\cite{serafini2017quantum,monras2013ARXIV}:
\begin{equation}\label{QFI_gaussian}
\mathcal{F}_{\theta}^{\;\text{Q}}(\theta)=\frac{\text{Tr}[(\boldsymbol{\sigma}^{-1}\partial_{\theta}\boldsymbol{\sigma})^{2}]}{2(1+\mu^{2})}
+2\frac{(\partial_{\theta}\mu)^2}{1-\mu^4}
+2(\partial_{\theta}\boldsymbol{d})^{\text{T}}(\boldsymbol{\sigma}^{-1})(\partial_{\theta}\boldsymbol{d}),
\end{equation}
where $\mu = 1/\sqrt{\det(\boldsymbol{\sigma})}$ is the purity of the state, and $\partial_{\theta}$ denotes differentiation with respect to the parameter $\theta$. The three terms represent contributions from the parameter-dependence of the covariance matrix, the purity, and the displacement vector, respectively.

% The \nocite command causes all entries in a bibliography to be printed out
% whether or not they are actually referenced in the text. This is appropriate
% for the sample file to show the different styles of references, but authors
% most likely will not want to use it.
\nocite{*}

\bibliography{reference}% Produces the bibliography via BibTeX.

\end{document}